\documentclass[epjc3]{svjour3}

\RequirePackage{fix-cm}
\RequirePackage{lineno}

\usepackage{graphicx}
\usepackage{caption}    
\usepackage{subfig}
\usepackage{amsmath}
\usepackage{array}
\usepackage{upgreek}
\usepackage{hyperref}
\usepackage{multirow}
\usepackage[numbers]{natbib}

\journalname{Eur. Phys. J. C}

\begin{document}

\title{Reconstruction of interactions in the ProtoDUNE-SP detector with Pandora}

%

\author{The DUNE Collaboration\\\\
       A.~Abed Abud\thanksref{Liverpool,CERN}
       \and B.~Abi\thanksref{Oxford}
       \and R.~Acciarri\thanksref{Fermi}
       \and M.~A.~Acero\thanksref{Atlantico}
       \and M.~R.~Adames\thanksref{Tecnologica }
       \and G.~Adamov\thanksref{Georgian}
       \and M.~Adamowski\thanksref{Fermi}
       \and D.~Adams\thanksref{Brookhaven}
       \and M.~Adinolfi\thanksref{Bristol}
       \and C.~Adriano\thanksref{Campinas}
% ...
}

\institute{
% ...
\pagebreak[0] Universidad del Atl\'antico, Barranquilla, Atl\'antico, Colombia\label{Atlantico}
% ...
\and\pagebreak[0] University of Bristol, Bristol BS8 1TL, United Kingdom\label{Bristol}
        \and\pagebreak[0] Brookhaven National Laboratory, Upton, NY 11973, USA\label{Brookhaven}
% ...
\and\pagebreak[0] Universidade Estadual de Campinas, Campinas - SP, 13083-970, Brazil\label{Campinas}
% ...
\and\pagebreak[0] CERN, The European Organization for Nuclear Research, 1211 Meyrin, Switzerland\label{CERN}
% ...
\and\pagebreak[0] Fermi National Accelerator Laboratory, Batavia, IL 60510, USA\label{Fermi}
% ...
\and\pagebreak[0] Georgian Technical University, Tbilisi, Georgia\label{Georgian}
% ...
\and\pagebreak[0] University of Liverpool, L69 7ZE, Liverpool, United Kingdom\label{Liverpool}
% ...
\and\pagebreak[0] University of Oxford, Oxford, OX1 3RH, United Kingdom\label{Oxford}
% ...
\and\pagebreak[0] Universidade Tecnol\'ogica Federal do Paran\'a, Curitiba, Brazil\label{Tecnologica }
% ...
}

\titlerunning{The Pandora Reconstruction for ProtoDUNE-SP}

\date{Received: date / Accepted: date}

\maketitle

\begin{abstract}
The Pandora Software Development Kit and algorithm libraries provide pattern-recognition logic essential to the reconstruction of particle interactions in liquid argon time projection chamber detectors. Pandora is the primary event reconstruction software used at ProtoDUNE-SP, a prototype for the Deep Underground Neutrino Experiment far detector. ProtoDUNE-SP, located at CERN, is exposed to a charged-particle test beam. This paper gives an overview of the Pandora reconstruction algorithms and how they have been tailored for use at ProtoDUNE-SP. In complex events with numerous cosmic-ray and beam background particles, the simulated reconstruction and identification efficiency for triggered test-beam particles is above 80\% for the majority of particle type and beam momentum combinations. Specifically, simulated 1\,GeV/$c$ charged pions and protons are correctly reconstructed and identified with efficiencies of 86.1$\pm0.6$\% and 84.1$\pm0.6$\%, respectively. The efficiencies measured for test-beam data are shown to be within 5\% of those predicted by the simulation.
\end{abstract}

\section{Introduction}
\label{sec:intro}
ProtoDUNE-SP~\cite{protoduneDetector} was a single-phase (SP) liquid argon time projection chamber (LArTPC) detector prototype for the Deep Underground Neutrino Experiment (DUNE) far detector~\cite{dunetdr_v1}. Installation of the detector at the CERN Neutrino Platform was completed in August 2018, and charged-particle test-beam data were collected from August 2018 until the start of the CERN long shutdown period in December 2018.

The primary engineering goal of the ProtoDUNE-SP detector was to prototype the production of large-scale LArTPCs for use at the DUNE far detector (FD)~\cite{dunetdr_v4}. Alongside the validation of production and installation procedures, ProtoDUNE-SP had goals related to testing the event reconstruction and performing detector calibration in a controlled environment. The primary physics goals were measurements of the interaction cross-sections for various charged particle species on a liquid argon target that will be very valuable for modelling neutrino interactions at DUNE.

Pandora is a software package that has been developed for event reconstruction in high energy physics and is now in use at ProtoDUNE-SP~\cite{proto_performance}. It consists of a framework, the Pandora Software Development Kit (SDK)~\cite{pandorasdk}, and a number of experiment-specific content libraries containing pattern-recognition logic. Originally developed for event reconstruction at future linear $e^{+}e^{-}$ colliders~\cite{pandorailc,pandoraclic}, Pandora has since been successfully applied in LArTPC experiments, such as MicroBooNE~\cite{pandorauboone}. Pandora brings a multi-algorithm philosophy to LArTPC event reconstruction, applying over 100 algorithms to develop the reconstruction from the input hits to a hierarchy of fully-reconstructed particles. Each algorithm is designed to address a specific aspect of event reconstruction, and they collectively provide robust and sophisticated pattern recognition. Pandora incorporates machine-learning techniques, such as boosted decision trees (BDTs)~\cite{bdtOriginal,bdtBoost} and support vector machines~\cite{cortes1995support}, to drive decisions made at certain junctions of the event reconstruction. The Pandora event reconstruction can be run standalone and it has also been interfaced with LArSoft~\cite{Church:2013hea}, a common software framework used by the majority of LArTPC experiments.

The contents of this paper are as follows: Sec.~\ref{sec:expdets} describes the ProtoDUNE-SP experiment, Sec.~\ref{sec:patrec} describes the Pandora reconstruction, Sec.~\ref{sec:datadescription} describes the simulated and experimental data samples, Sec.~\ref{sec:cosmicrecoperformance} provides an assessment of the cosmic-ray reconstruction, Sec.~\ref{sec:testbeamrecoperformance} examines the performance of the test-beam reconstruction, and Sec.~\ref{sec:conclusion} provides concluding remarks.

\section{Experimental Details}
\label{sec:expdets}

\subsection{Charged Particle Test Beam}

A dedicated extension~\cite{PhysRevAccelBeams.20.111001,PhysRevAccelBeams.22.061003} to the CERN H4 beamline was constructed for ProtoDUNE-SP. The test beam contains a mixture of particle species: $\pi^{+}$, $e^{+}$, $p$, $\mu^{+}$, and $K^{+}$. The polarity of the beam focusing magnets can be reversed to produce a beam containing negatively charged particles, but all test-beam data collected in 2018 was taken in the positive polarity mode. The beam momentum can be varied from 0.3 to 7\,GeV/$c$ with a resolution of $\Delta p/p \leq 3\%$~\cite{pdtdr}, providing particles with similar energies as those expected to be produced in the 0.5 to 5.0\,GeV neutrino interactions in the DUNE FD~\cite{dune_sense}. The test beam enters the detector through a beam plug in the upstream face and is approximately 10\,cm in diameter. The beam line has numerous instruments that are used to trigger the detector readout electronics, to measure the momentum and the trajectory of the test-beam particles prior to their entrance into the detector, and to identify their species. Full details of the test-beam design can be found in Refs~\cite{proto_performance,PhysRevAccelBeams.20.111001,PhysRevAccelBeams.22.061003}.

\subsection{ProtoDUNE-SP}
\label{sec:protodunesp}
\begin{figure}[htb]
    \centering
    \includegraphics[width=0.6\textwidth]{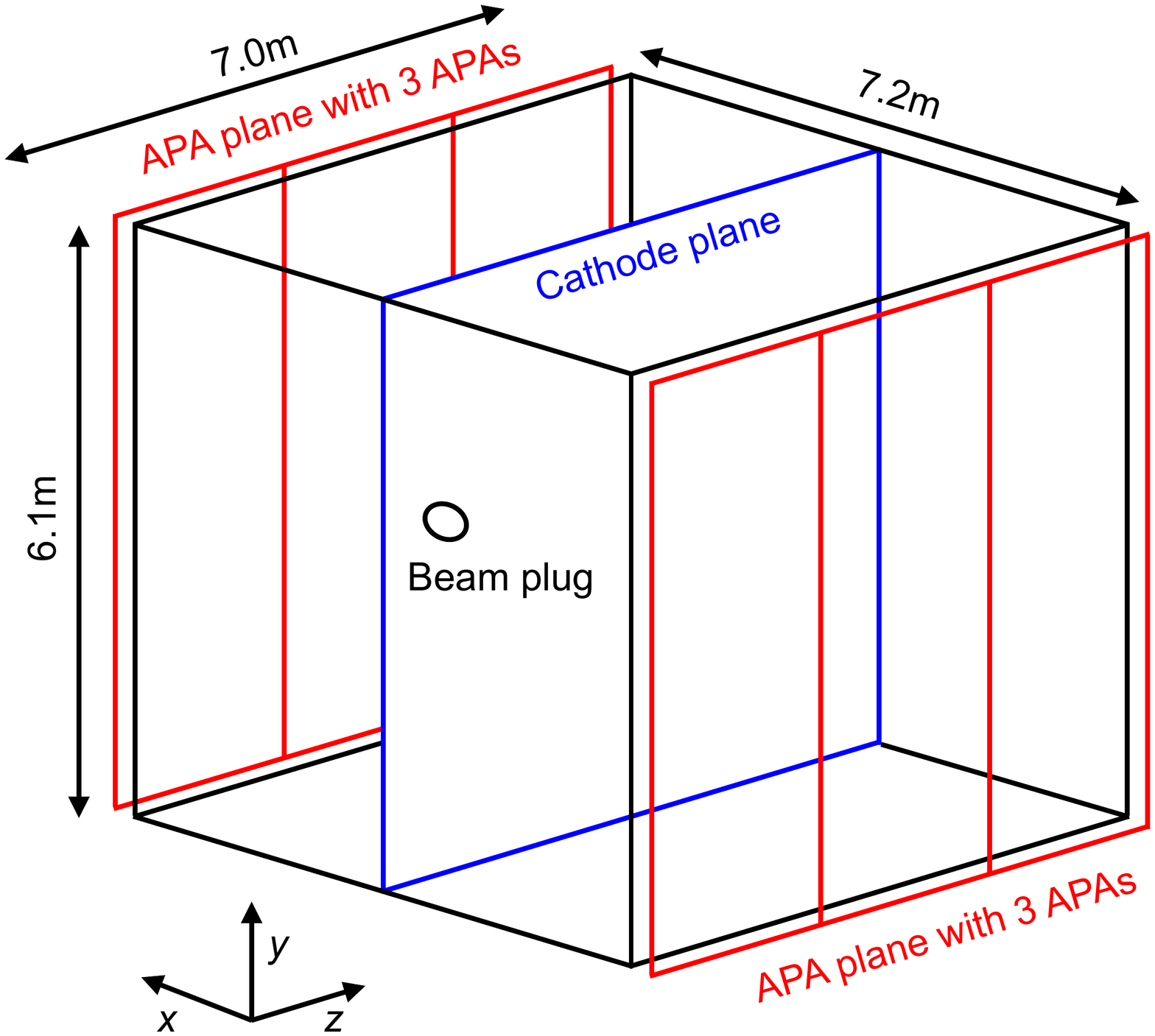}
    \includegraphics[width=0.3\textwidth]{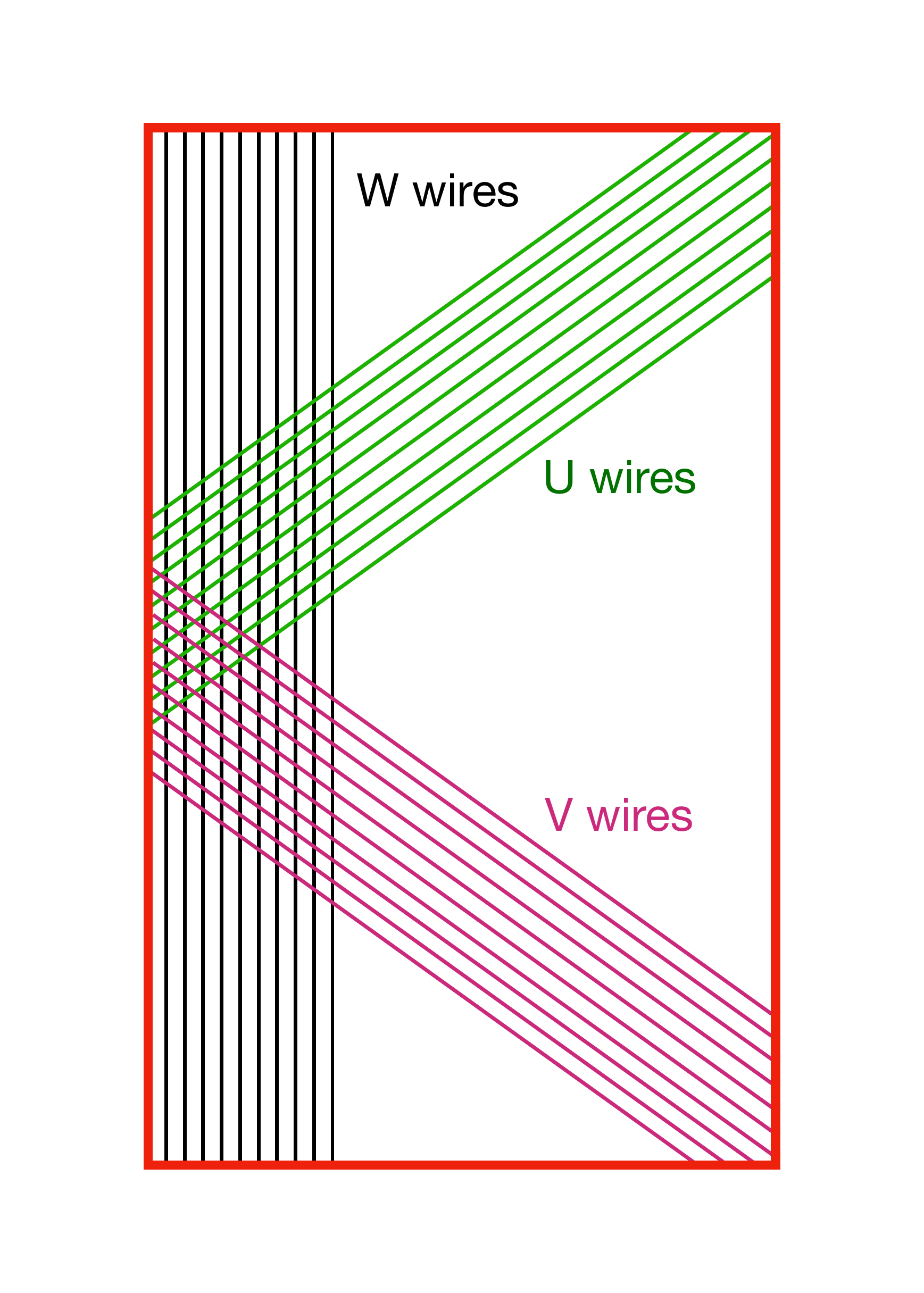}
    \caption{Left: a simple drawing of the ProtoDUNE-SP detector. The black box represents the active volume, divided into two parts by the central cathode (blue). The six APAs are arranged into two planes (red) either side of the cathode. The test beam enters through the beam plug, close to the right side of the cathode. The right-handed coordinate system is shown in addition to the dimensions of the active volume. Right: an illustration of the three wire planes on an APA, with only ten wires for each plane shown for clarity.}
    \label{fig:schematic3D}
\end{figure}

The ProtoDUNE-SP detector is extensively described in Refs~\cite{protoduneDetector} and~\cite{proto_performance}. A simplified schematic of the detector is shown in Fig.~\ref{fig:schematic3D}. It has a cuboid geometry with active-volume dimensions: 7.2\,m (width), 6.1\,m (height) and 7.0\,m (length). It has a total liquid argon mass of 0.77\,kt making it the largest LArTPC constructed to date\footnote{The dual-phase LArTPC ProtoDUNE-DP, which was built shortly after ProtoDUNE-SP, was approximately the same size.}. The nominal electric field in the active volume is 500\,V/cm, generated by the cathode plane (an array of Cathode Plane Assembly (CPA) modules), which is held at -180\,kV, and two sets of three Anode Plane Assemblies (APAs), one on either side of the central cathode, which are effectively grounded. The field cage ensures the uniformity of the electric field and shields it from the cryostat walls. The APAs are two-sided such that they can read out a drift volume on either side, as required for the DUNE FD. Each side of the APA has a plane of collection wires, referred to as the \textit{w} planes, that collect the ionisation charges from that side of the APA. In front of the collection plane there are two planes of induction wires, the inner one is denoted $v$ and the outer $u$, that wrap around both sides of the APA. The \textit{w} plane wires are vertical with a 4.790\,mm pitch between the wires. The \textit{u} and \textit{v} plane wires are aligned at $\pm35.7^\circ$ to the vertical, with a pitch of 4.669\,mm between the wires.

A right-handed Cartesian coordinate system is used to describe the detector geometry: $x$ defines the drift axis and is either equal to or opposite to the drift direction, $y$ is the upwards vertical direction and $z$ the remaining orthogonal direction. The test beam is directed primarily along the positive $z$ direction. The origin of the coordinate system is at the bottom of the upstream end of the cathode.

Each wire measures the induced or collected charge as a function of time for the duration of each 3\,ms readout window. The maximal drift time, defined as the time taken for charge to drift from the cathode to the APAs, is approximately 2250\,$\upmu$s. The 3\,ms readout window, relative to the beam trigger time, spans the range from $-$250\,$\upmu$s to 2750\,$\upmu$s and ensures that any charge deposited in the detector at the beam trigger time will be collected. Various detector effects are removed to reduce noise from the raw waveforms. This process consists of the mitigation of noisy readout wires, the removal of coherent and high-frequency noise from the wire signals and the deconvolution of the wire signals. The deconvolution procedure identifies and accounts for signals on a given wire that are produced via induction when adjacent wires observe charge~\cite{wirecell,uBooNESignal}. Hits are then formed from the collected or induced charge waveforms by fitting Gaussian functions to peaks in the waveforms. The wrapping of the induction wires requires that a disambiguation procedure using time-based coincidences between the induction views and the collection view is used to eliminate \textit{ghost} hits. Full details of the disambiguation procedure are given in Ref.~\cite{dunetdr_v2}. Each wire plane yields a 2D view of particle interactions in the LArTPC that forms the input to the pattern-recognition algorithms. A full description of signal processing, noise removal and charge calibration is given in Ref.~\cite{proto_performance}.

\begin{figure*}[htb]
\centering
\includegraphics[width=0.8\textwidth,trim={0 0 4.6cm 0},clip]{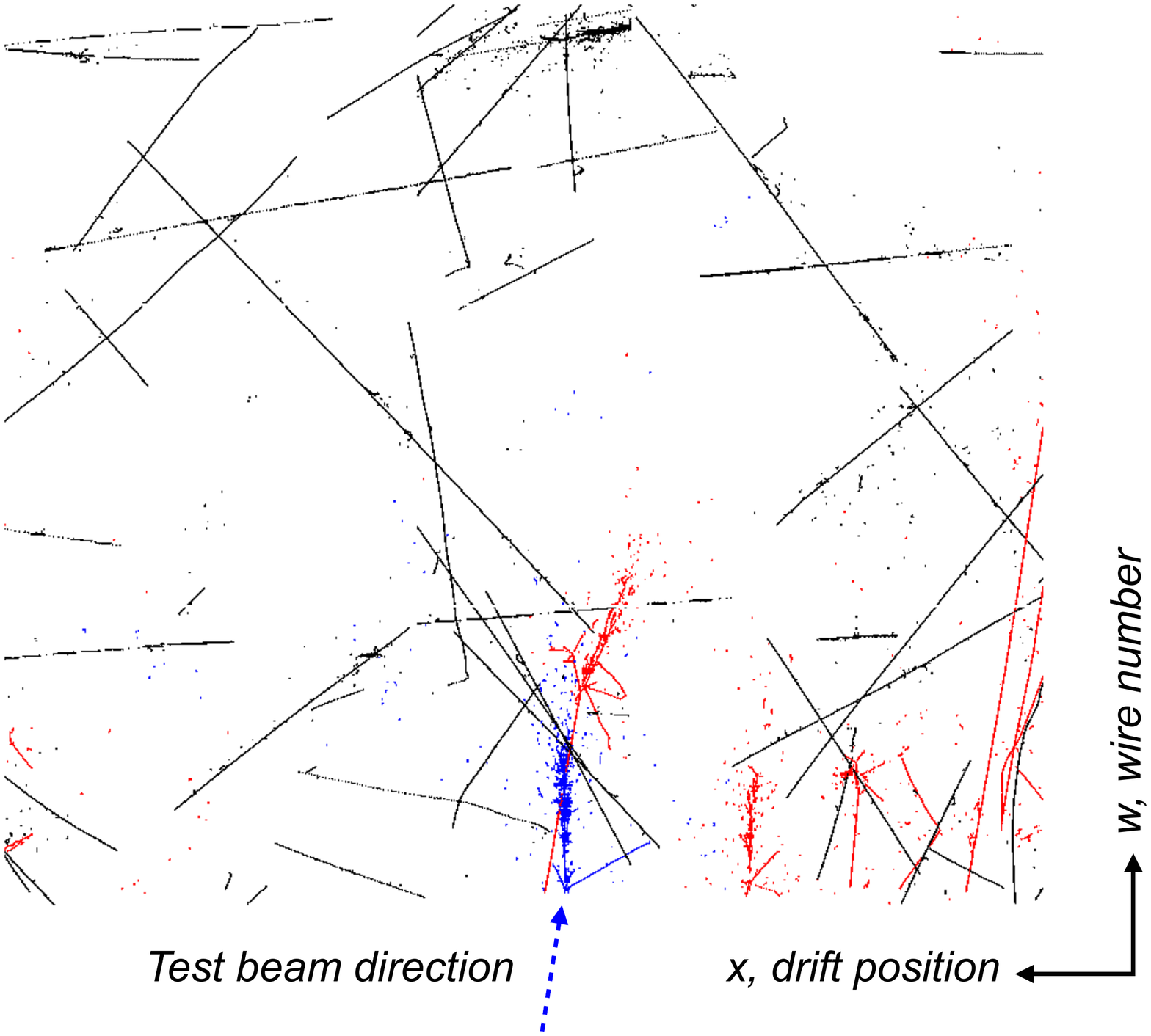}
\caption{An example of the $w$ view for a simulated 7\,GeV/$c$ $\pi^{+}$ event at ProtoDUNE-SP where the hits have been coloured to indicate their origin: triggered test-beam particle (blue), beam halo (red) and cosmic-ray muon (black). The vertical axis is equivalent to the $z$ axis of the detector and the horizontal axis is converted from the drift time. The test beam particles enter from the upstream end (bottom of the image).}
\label{fig:eventdecomp}
\end{figure*}

Figure~\ref{fig:eventdecomp} shows the reconstructed hits for a typical simulated 3\,ms readout window in ProtoDUNE-SP in the collection ($w$) view in the (drift coordinate, wire position) parameter space. For the $w$ view, the wire position is the same as the $z$ coordinate, and the drift time has been converted to the spatial $x$ coordinate using the drift velocity. The three colours represent hits from three classes of particles (and any subsequent particles produced in their interactions): blue shows the \textit{triggered test-beam} 7\,GeV/$c$ charged pion particle that initiated the readout of the detector and interacted almost immediately after entering the TPC; red shows all other beam particles, henceforth called \textit{beam halo} particles, which includes particles from interactions in the beam line, those not focused by the beam-line magnets due to their momentum, particle decays, and focused particles that arrived within the readout window of the triggered beam particle; and black shows cosmic rays (mostly muons). 

In general, test-beam particle interactions produce complex particle hierarchies (containing secondary, tertiary, etc., particles from interactions and decays) involving both track-like and shower-like energy deposits, while cosmic-ray muons primarily produce track-like topologies. Due to the surface location of ProtoDUNE-SP, the majority of the collected charge signals originated from cosmic-ray muons that traverse the detector throughout the 3\,ms readout window. The measured time of charge collected on the wires, $t_m$, is a function of the time that the particle entered the detector ($t_0$) relative to the time that the detector readout was triggered ($t_\textrm{trigger}$), and the distance in the drift coordinate from the APA to where the energy was deposited ($x$):
\begin{linenomath*}
\begin{equation}
\label{eq:ambiguity}
    t_m = t_0 - t_\textrm{trigger} + x/v_d,
\end{equation}
\end{linenomath*}
where $v_d$ is the electron drift velocity. By definition in ProtoDUNE-SP, the trigger time $t_\textrm{trigger} = 0$ and at the nominal electric field the electron drift velocity is $v_d$ = 1.59\,mm/$\upmu$s~\cite{proto_performance}.

\begin{figure}[htb]
\centering
\includegraphics[width=0.6\textwidth]{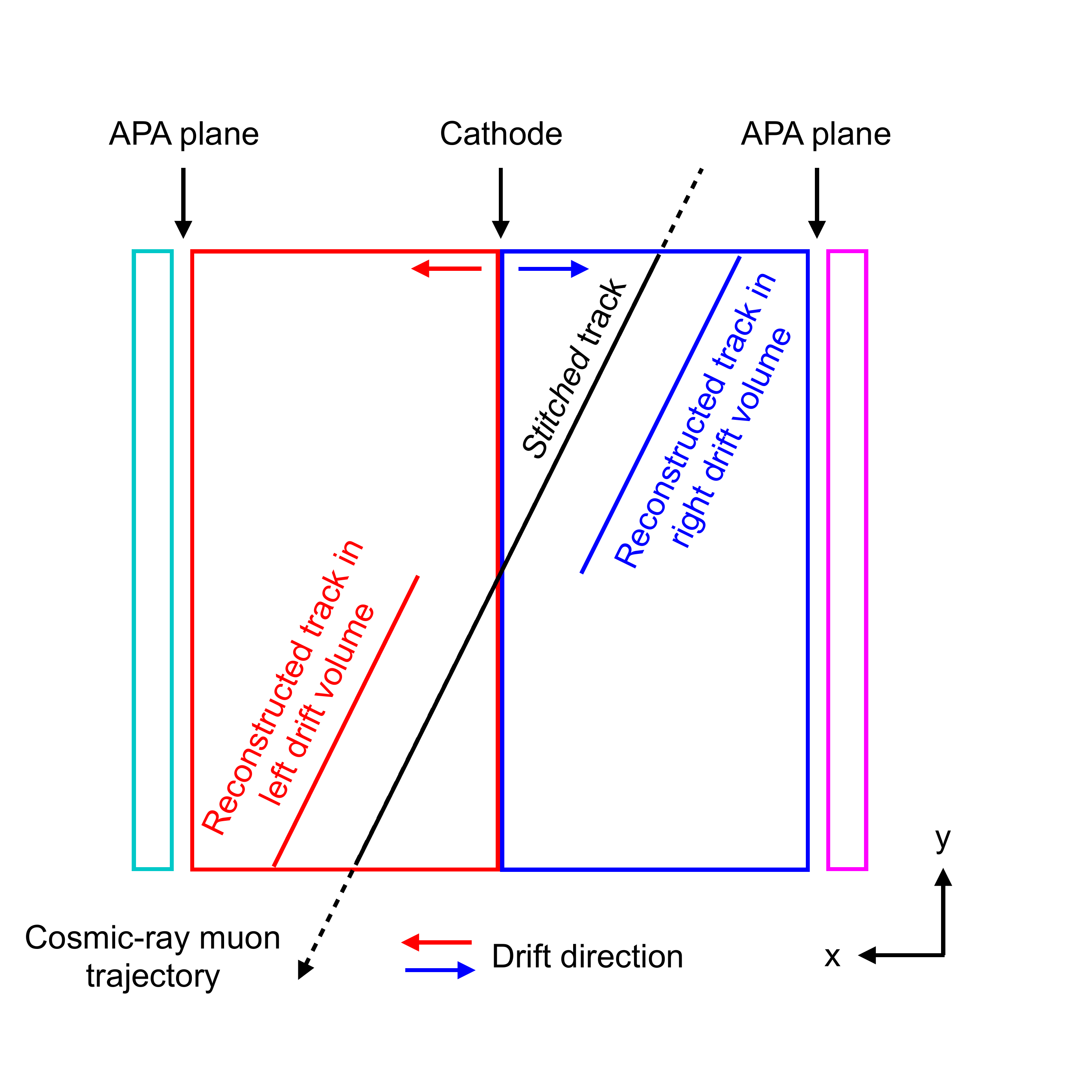}
\caption{An example of a cosmic ray crossing the detector from top to bottom and passing through the cathode. Under the initial (and incorrect) assumption of $t_0=0$ the energy depositions in the two drift volumes (red and blue lines) appear to be at the wrong position in the drift direction. The reconstruction can recover the correct $t_0$ by \textit{stitching} the two tracks at the cathode by shifting the drift coordinate in each drift volume by an equal and opposite amount, resolving the ambiguity in the drift coordinate position.}
\label{fig:stitching}
\end{figure}

There is an ambiguity in Equation~\ref{eq:ambiguity} between $t_0$ and $x$ for a given $t_m$ unless $t_0$ is known. For the test-beam particle that triggers the TPC readout $t_0 = t_\textrm{trigger}$ by definition and there is no ambiguity. In all other cases, the $t_0$ for any particle is initially undetermined and is assumed to arrive at the beam trigger time and hence assigned a preliminary value of $t_0=0$. The exact position in the drift coordinate inside the TPC where charge was physically deposited is thus only well known for triggered test-beam particles. For other particles (cosmic rays and beam-halo particles) this intrinsic ambiguity in the drift coordinate makes it initially impossible to distinguish between charge deposited by a particle arriving before the beam trigger but far from the APA, and charge deposited by a particle arriving after the beam trigger but close to the APA, since in both cases the time at which the charge is collected by the readout wires would be the same. Figure~\ref{fig:stitching} shows an example of a cosmic ray with $t_0\neq0$ where the hits (red and blue lines) appear to be in the wrong position in the drift direction. For this event, the reconstruction can use the fact that the cosmic ray crossed the cathode to measure the correct $t_0$ and resolve the ambiguity, using the \textit{stitching} process explained in Sec.~\ref{sec:pandoracosmic}. Other LArTPCs have demonstrated that this ambiguity can be resolved for some interactions by matching the charge information with precise timing information from a photon detector system~\cite{ubLight1,ubLight2}.

\begin{figure}[htb]
    \centering
    \includegraphics[width=0.48\textwidth]{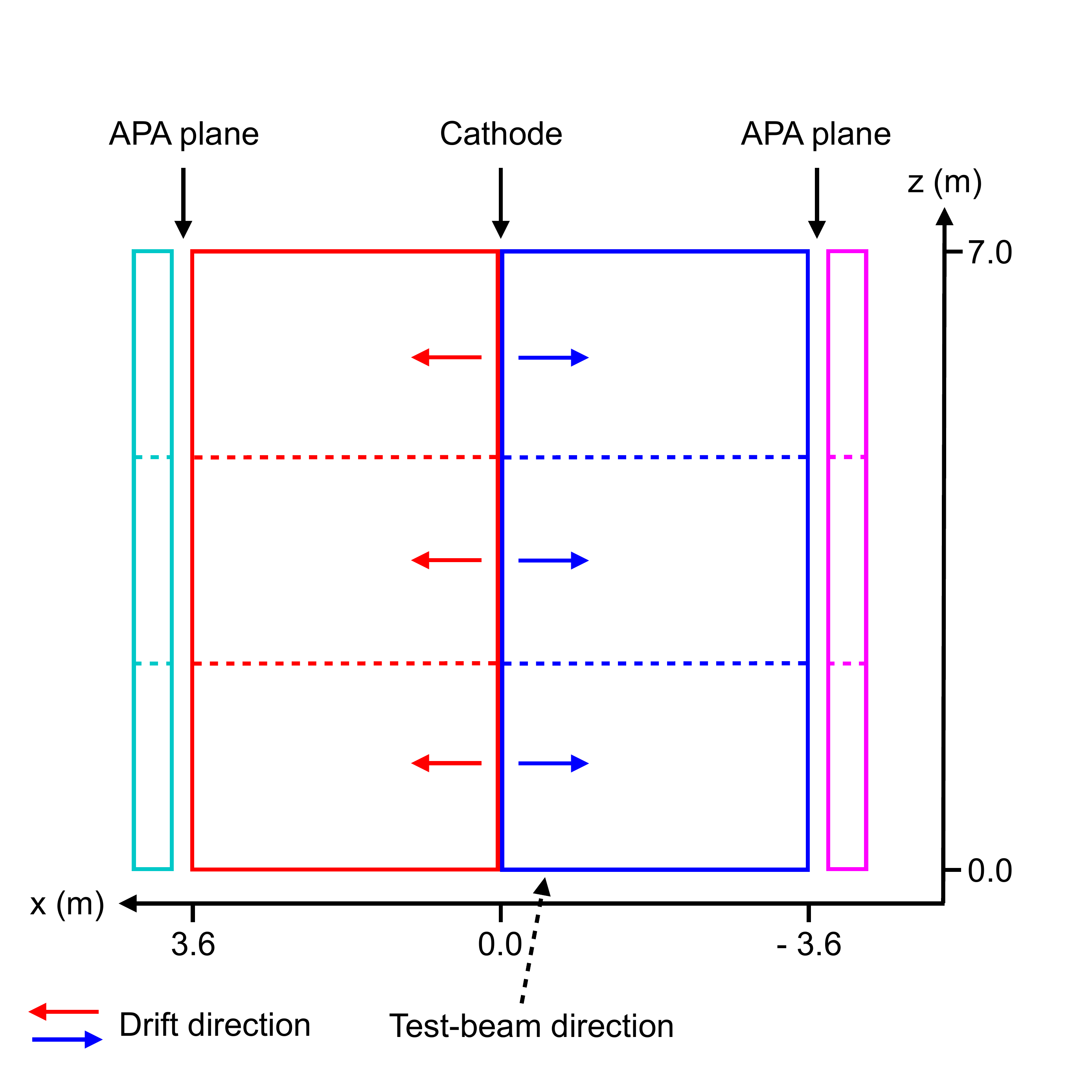}
    \caption{A schematic top-down view of ProtoDUNE-SP. The six drift regions, each read out by an individual APA, are separated by the dashed lines within the larger drift volumes (red and blue) used inside Pandora. The dummy regions are also combined into dummy volumes (cyan and magenta). The test beam enters the detector close to the cathode and at $z=0$.}
    \label{fig:tpcs}
\end{figure}

The six APAs are read out independently and give rise to six volumes with drift fields (henceforth referred to as \textit{drift regions}), three on either side of the cathode. Since the APAs are read out on both sides, there are also six small volumes without drift fields between the APAs and the cryostat wall (known as \textit{dummy regions}), where charge can be detected if a particle crosses the APA. Figure~\ref{fig:tpcs} shows how adjacent drift regions (separated by the dashed lines) sharing a common drift direction are concatenated together inside Pandora to form two \textit{drift volumes} (red and blue), and the two sets of dummy regions are also concatenated into \textit{dummy volumes} (cyan and magenta). The drift direction for a given drift region is either along positive or negative $x$ depending on the local cathode and APA orientation. The merged drift volumes allow the pattern recognition to trivially correlate inputs between adjacent (in the $z$ direction) drift regions. 

\subsection{Space Charge Effects}
\begin{figure}[htb]
\centering
\includegraphics[width=0.6\textwidth]{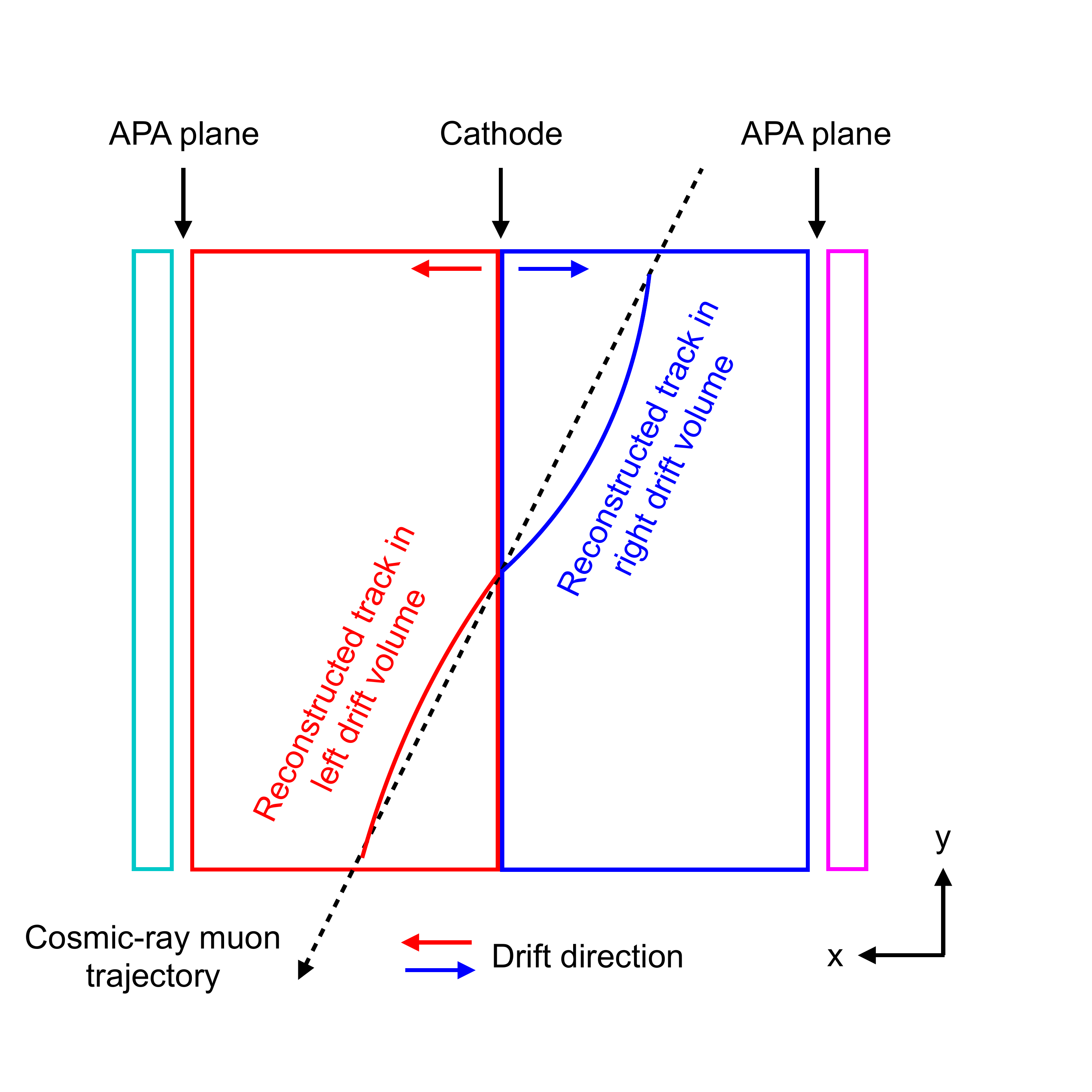}
\caption{A schematic diagram of how the space charge effect causes distortions to the reconstructed tracks in both the drift direction and the orthogonal directions. For clarity, the amount of distortion has been exaggerated. The red and blue show the tracks reconstructed in the two central drift volumes, while the black dotted line represents the true trajectory of the cosmic-ray muon.}
\label{fig:spacechargetrack}
\end{figure}

Surface-based LArTPC detectors such as ProtoDUNE-SP are subject to space charge effects (SCE): the build up of slow-moving\footnote{Approximately $5\times10^{-5}$mm/$\upmu$s~\cite{ionDrift}} positive ions in the detector due to the high rate of cosmic-ray muons. These ions are produced when charged particles pass through the detector and ionise the liquid argon~\cite{spacecharge}. This effect distorts the electric field by up to 25\% in some regions of the detector and hence the drift velocity of ionisation electrons within the LArTPC. Assuming the nominal uniform electric field, the positions of the reconstructed hits within the detector are shifted with respect to their true positions, and the charges of the hits are also distorted. The simulation uses a data-driven space charge distortion based on measurements of cosmic-ray muons, as described in Ref.~\cite{proto_performance}, which is asymmetric with respect to the cathode. Figure~\ref{fig:spacechargetrack} illustrates how the observed tracks in the two drift volumes show a characteristic bowing effect in the drift direction compared to the true straight trajectory. While the SCE is an important consideration for surface detectors, its impact will be much smaller for deep underground detectors such as the DUNE far detectors due to the significantly lower rate of cosmic rays.

\section{Pandora Event Reconstruction}
\label{sec:patrec}

The Pandora event reconstruction for ProtoDUNE-SP builds directly upon the approaches and algorithms developed for MicroBooNE, described in Ref.~\cite{pandorauboone}. In this section, the emphasis is on how these algorithms have been extended and harnessed to reconstruct cosmic-ray muon and test-beam particles in a surface-based LArTPC detector comprising multiple drift volumes.  

The inputs to the Pandora pattern recognition are hits, and each hit represents a signal detected on a specific wire at a specific time. The input hits each have a drift time coordinate and a wire coordinate, and are associated with a specific readout plane. Hence, three 2D views are presented as inputs: the $u$, $v$ and $w$ views. For ProtoDUNE-SP, one additional piece of information is collected per hit: an index identifying the drift volume from which the hit originates.

Pandora uses a multi-algorithm approach to pattern recognition, and the three 2D inputs are examined by a series of over one hundred algorithms and tools, which gradually identify features and build up a picture of events. The final goal is for each true or real particle to be reconstructed as a single reconstructed particle, that is both pure (containing only hits from that particle) and complete (containing all hits from that particle). The overall approach is to:

\begin{figure*}[htb]
    \centering
    \includegraphics[width=0.9\textwidth]{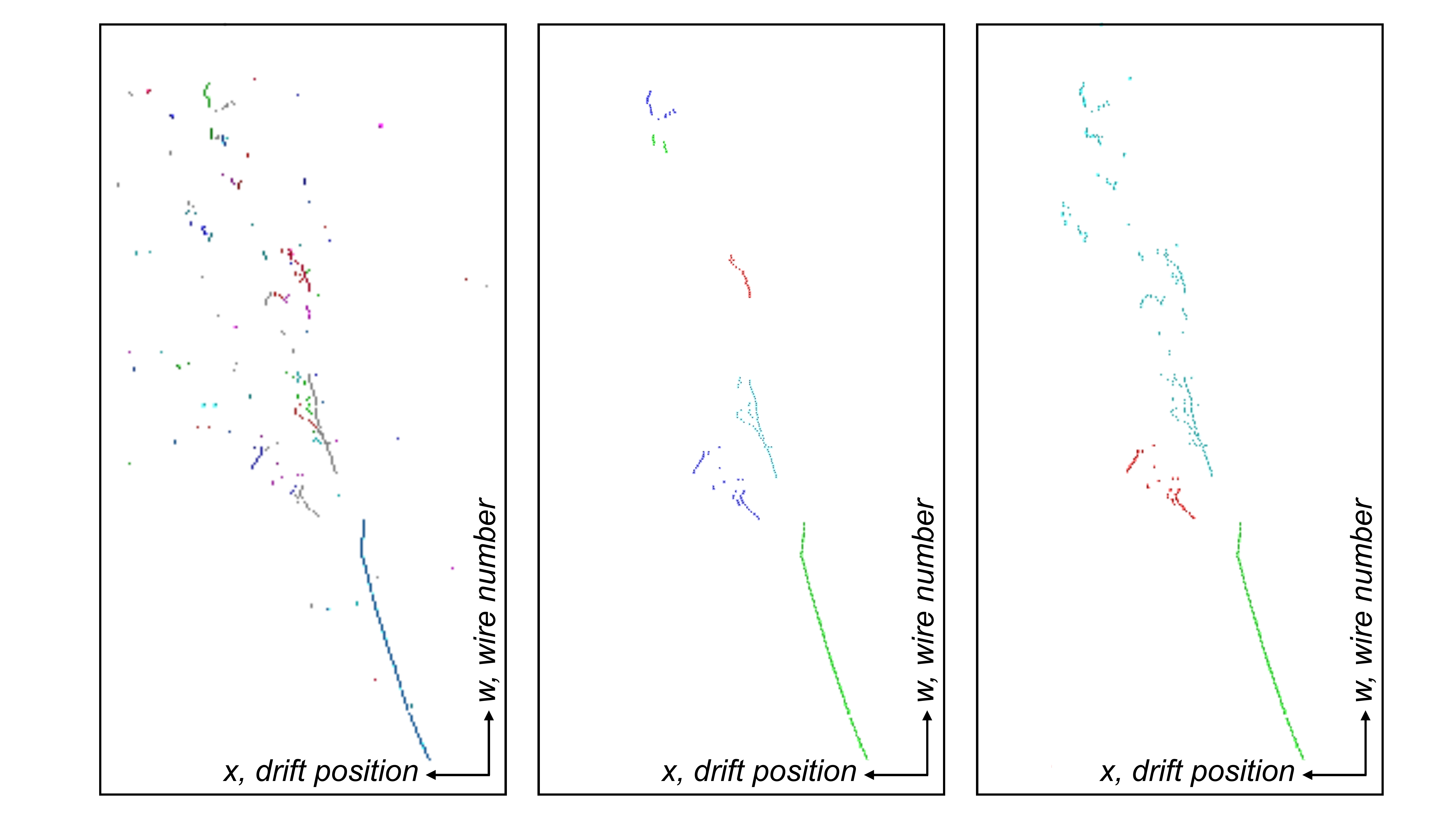}
    \caption{A ProtoDUNE-SP data beam interaction shown at different stages of the reconstruction: after initial clustering (left), after particle creation (middle), and after full hierarchy reconstruction (right). All images show the reconstructed hits in the collection (\textit{w}) view.}
    \label{fig:clustering}
\end{figure*}

\begin{itemize}
    \item \textbf{Assign hits to clusters.} The hits from each readout plane are considered separately and clustered, with the aim of creating one cluster per input particle. This procedure begins with a main clustering algorithm that is designed to be conservative to prioritise making pure but incomplete clusters and avoiding mistakes. A series of subsequent ``topological association'' algorithms exploit the detector granularity to increase the cluster completeness, whilst maintaining purity.
    \item \textbf{Assign clusters to particles.} The clusters from each readout plane are compared, by exploiting the drift coordinate common to all planes, and by using knowledge of the wire angles to correlate features in 3D. By comparing the independent 2D pattern-recognition outcomes for the three planes, corrections to the 2D clustering can be made, and clusters can be unambiguously associated between planes. Clusters are bound together to form reconstructed particles.
    \item \textbf{Assign particles to hierarchies.} Reconstructed particles contain clusters from two or more readout planes, allowing a list of 3D hits to be created for each. A series of further topological association algorithms (now operating in 3D) continue to grow particle completeness, without sacrificing purity. The reconstruction concludes by organising the final particles into a 3D hierarchy, representing the final state particles and any subsequent interactions or decays.
\end{itemize}
Figure~\ref{fig:clustering} shows the sequential progression of the reconstruction of a beam interaction following the three main stages outlined above.

Pandora had two chains of algorithms for event reconstruction in neutrino detectors, PandoraCosmic and PandoraNu, for reconstructing interactions under cosmic-ray and neutrino hypotheses, respectively. In this section, details are presented of how the PandoraCosmic chain has been adapted for a detector with multiple drift volumes, and how the PandoraNu chain has been adapted to represent the interactions of charged particles in a test beam (and renamed PandoraTestBeam). Finally, a description is provided of how these two algorithm chains are used together to provide a clear reconstruction output. The aim is to provide an unambiguous interpretation of the input hits at ProtoDUNE-SP as a list of identified cosmic-ray muon and test-beam particle hierarchies.

\subsection{PandoraCosmic}
\label{sec:pandoracosmic}

The PandoraCosmic algorithm chain was developed to reconstruct cosmic-ray muon trajectories. Following the initial track-like clustering and particle-creation algorithms, any remaining hits are used to seed and grow shower particles. Parent muon track particles are linked to child shower particles, representing Michel electrons or delta rays. The muons are assumed to be downward going, so their primary vertices are placed at their highest reconstructed $y$ coordinates.

Cosmic rays arrive throughout the detector readout window. As previously mentioned, all hits passed to the pattern recognition are placed assuming that they correspond to a particle arriving at $t_0=t_\textrm{trigger}$. For cosmic rays, this can result in a shift in the drift coordinate at which their hits are placed. This offset may place the hits outside of the physical drift volume, and this information can be used to tag cosmic rays.

The ProtoDUNE-SP detector has four adjacent volumes (the two central drift volumes and the two outer dummy volumes), which means cosmic-ray muons can cross between volumes, and this provides new information and a new challenge. The hits from one muon with $t_0\neq0$ will be shifted in each volume. But, as the drift direction alternates between adjacent volumes, their drift coordinates will be shifted in opposite directions. In the reconstruction, each drift volume is initially processed in isolation, resulting in separately reconstructed 3D particles in each volume. The separate particles are shifted by equal amounts in the drift direction, but the direction of the shift alternates between adjacent volumes. Shifting the particles in this way\footnote{There is some tolerance in these shifts between the two track segments to allow for an asymmetric SCE or slight imperfections in the track reconstruction.} should yield a single trajectory that is continuous in both its position and direction across the boundary between drift volumes. The separate component particles can then be \textit{stitched} together. The $t_0$ corrections identified by this stitching process allow for a single coherent 3D particle trajectory to be reconstructed, as demonstrated in Fig.~\ref{fig:stitching}. Performance metrics assessing the stitching performance and a discussion of how the SCE affects the measured $t_0$ are presented in Sec.~\ref{sec:crmetrics}.

\subsection{PandoraTestBeam}
\label{sec:pandoratestbeam}

The PandoraTestBeam algorithm chain is a modified version of the PandoraNu chain presented in Ref.~\cite{pandorauboone}. This chain focuses on identifying an interaction vertex, and reconstructing the individual track-like and shower-like particles that emerge from this point. Many of the algorithms are shared with the PandoraCosmic chain, but the vertex identification algorithms are specific to this chain, and there is a more sophisticated treatment of electromagnetic showers. The chain concludes with algorithms to build a hierarchy, representing the particle flow in the interaction.

\begin{figure}[tb]
\centering
\includegraphics[width=0.45\textwidth]{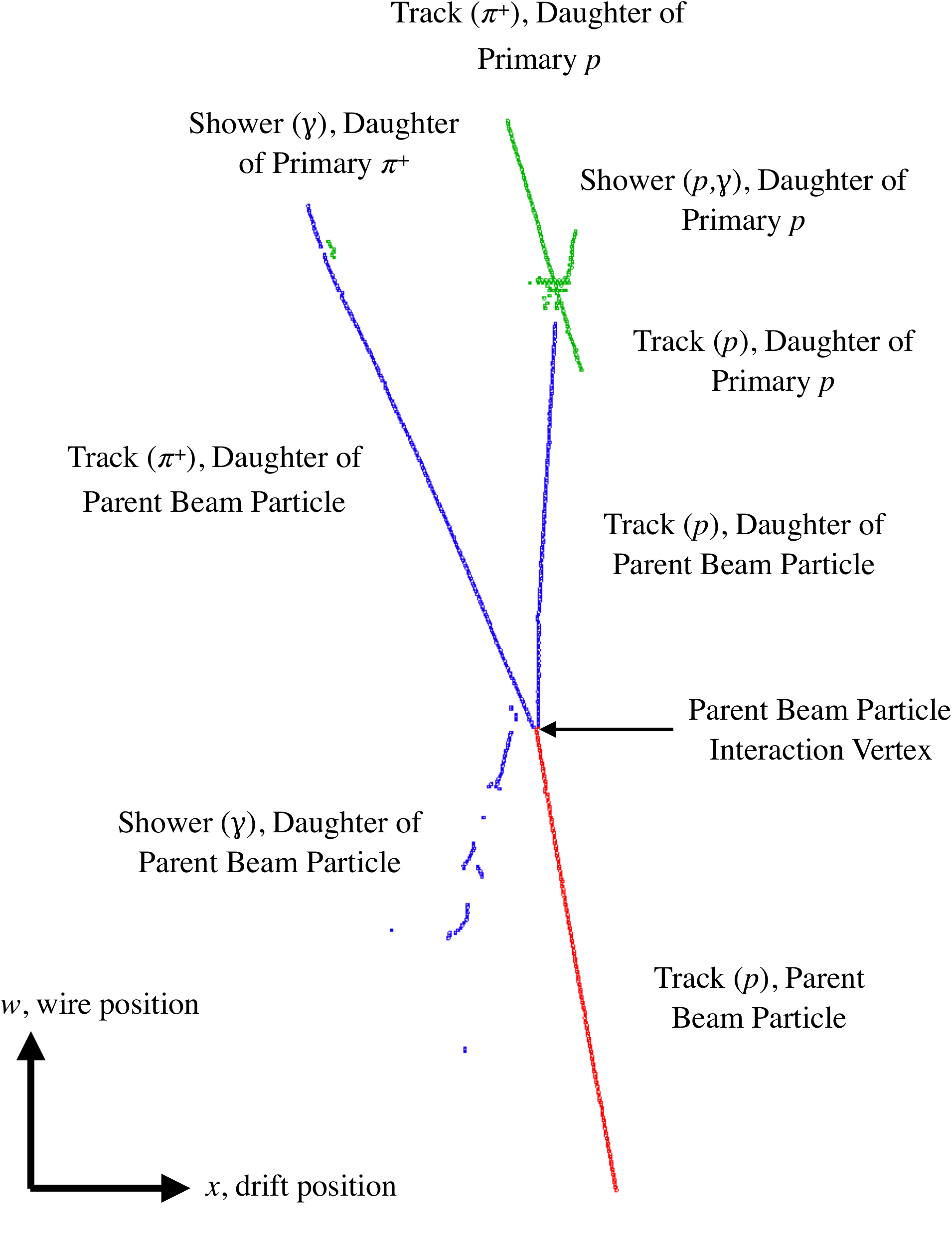}
\caption{An example of the 2D reconstruction output for a triggered test-beam particle. The particle hierarchy has been reconstructed to reflect the presence of an incoming track-like parent particle. The parent particle (red), child particles (blue) and subsequent child particles (green) have been separately highlighted.}
\label{fig:testbeamcreation}
\end{figure}

A new test-beam particle creation algorithm reorganises the hierarchy as appropriate for the interaction of an incoming charged test-beam particle. Particles initially reconstructed as emerging from the interaction vertex are reconsidered and the particle that is most consistent with actually being an \textit{incoming} particle is identified as the primary beam particle, which has both reconstructed start and interaction vertices. Parent-child links are formed between the primary beam particle and the other particles emanating from the interaction vertex to represent the newly-identified particle flow. Figure~\ref{fig:testbeamcreation} shows an example reconstructed particle hierarchy for a simulated test-beam proton interaction.

\subsection{Consolidated Reconstruction}
\label{sec:consolidatedreconstruction}

\begin{figure}[tb]
\centering
\includegraphics[width=0.45\textwidth]{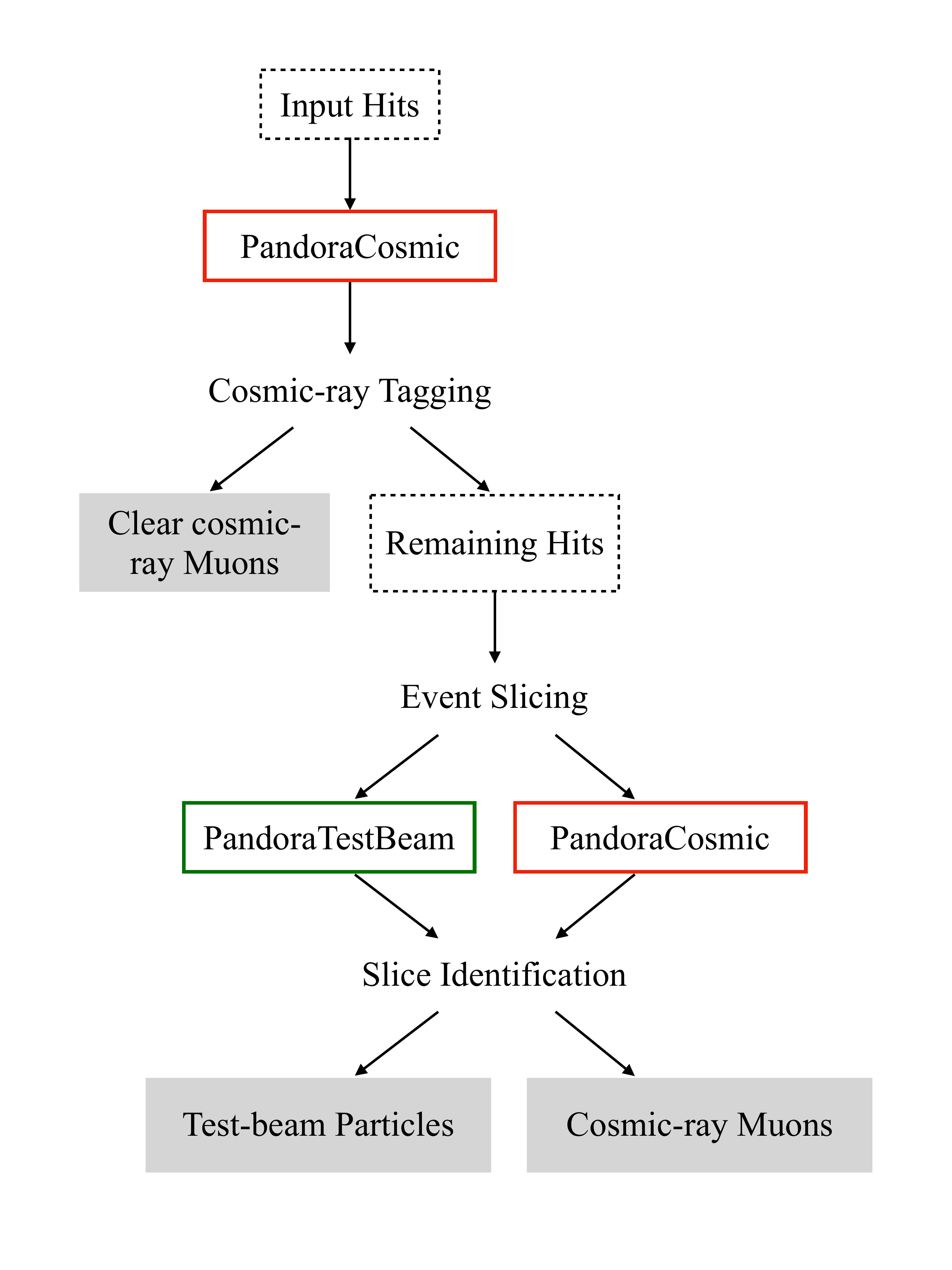}
\caption{Outline of the Pandora consolidated reconstruction. The PandoraTestBeam and PandoraCosmic algorithm chains run on the same hits in a given slice (a region of the detector containing hits originating from a single parent particle interaction) and yield two reconstruction outputs that can be compared and the optimal reconstruction selected.}
\label{fig:consolidatedreco}
\end{figure}

The aim of the Pandora consolidated reconstruction approach is to have one process that uses the PandoraCosmic and PandoraTestBeam algorithm chains to provide a clear and easy-to-interpret reconstruction output, with no double counting of any input hits. The output is a number of tagged reconstructed cosmic-ray particle hierarchies and tagged reconstructed test-beam particle hierarchies. A flow diagram illustrating the consolidated reconstruction approach is shown in Fig.~\ref{fig:consolidatedreco} and the following sections describe the different parts of this combined workflow: an initial pass of the PandoraCosmic chain, tagging of clear cosmic-ray muons, event slicing, parallel reconstruction of the slices using PandoraTestBeam and PandoraCosmic, and slice identification.

\subsubsection{Initial Pass of PandoraCosmic}

In the first step, hits from the four drift volumes are processed separately using the PandoraCosmic chain. The reconstructed particles from each drift volume then pass through the track stitching algorithm to fully reconstruct those particles that traverse neighbouring drift volumes. The reconstructed particle hierarchies are then passed to the cosmic-ray tagging algorithm.

\subsubsection{Cosmic-Ray Tagging}
\label{sec:stitching}

\begin{figure*}[htb]
\centering
\includegraphics[width=0.42\textwidth,trim={5.4cm 1cm 5cm 2cm},clip]{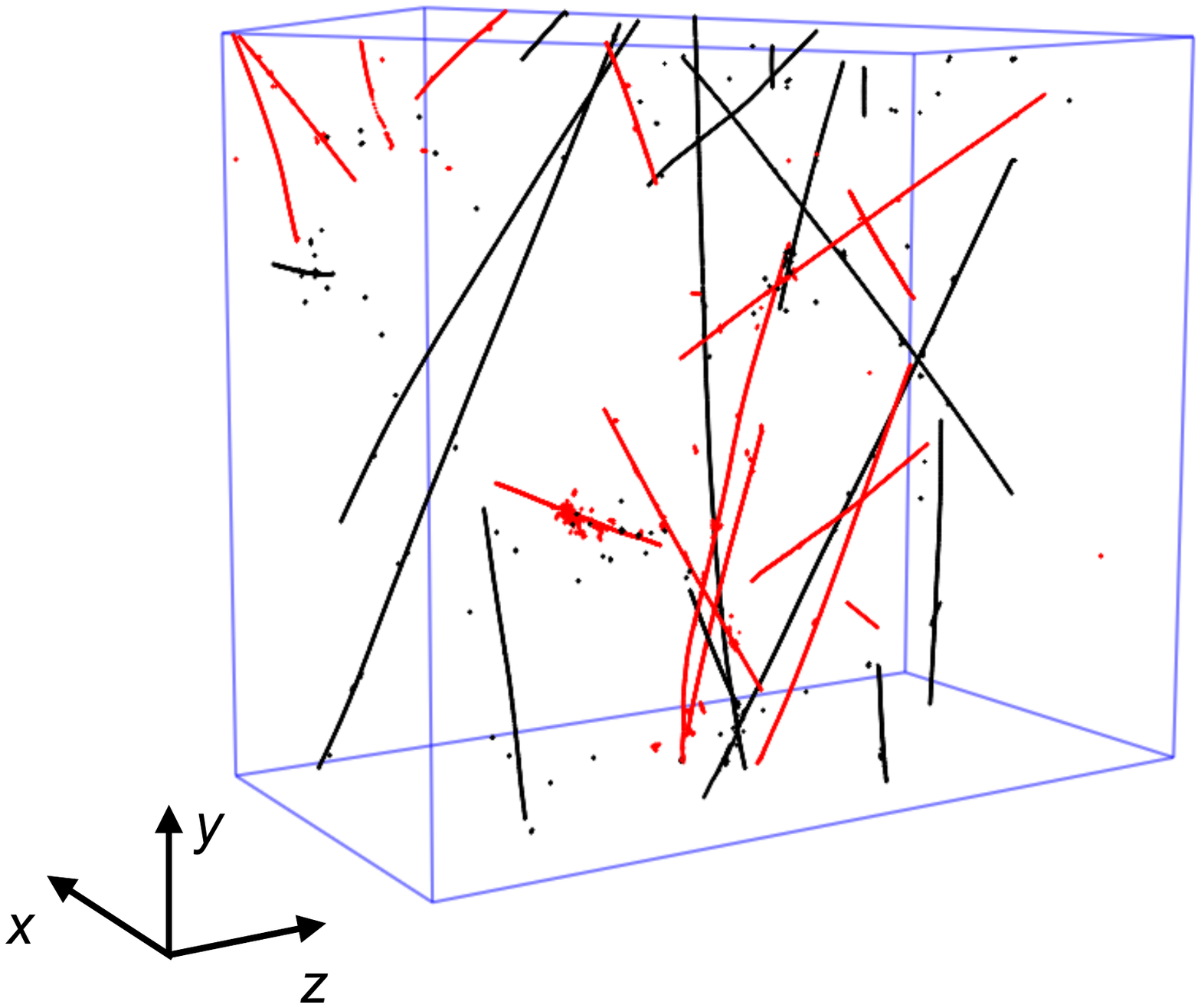}
\includegraphics[width=0.28\textwidth,trim={9cm 1cm 7cm 2cm},clip]{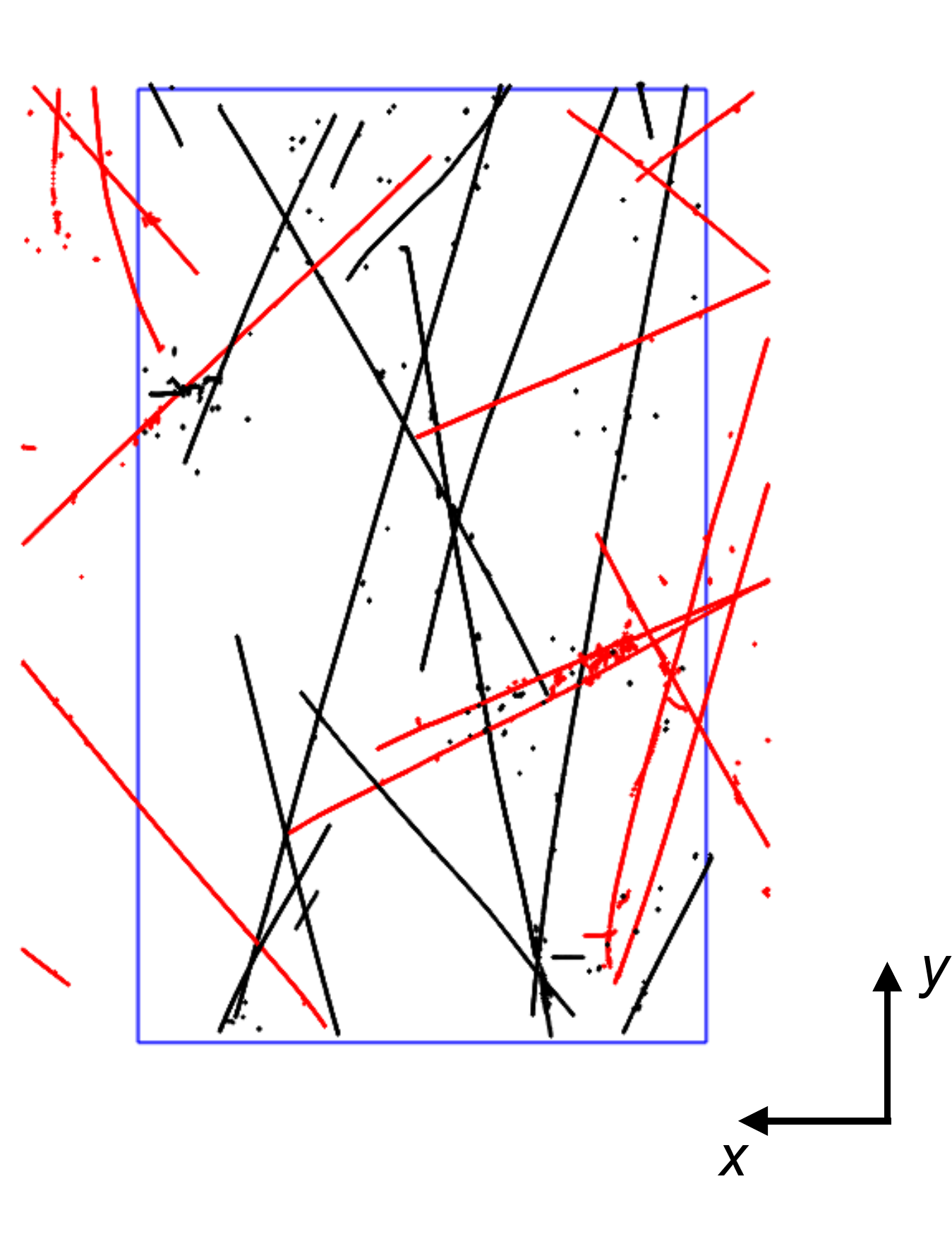}
\includegraphics[width=0.28\textwidth,trim={9cm 1cm 7cm 2cm},clip]{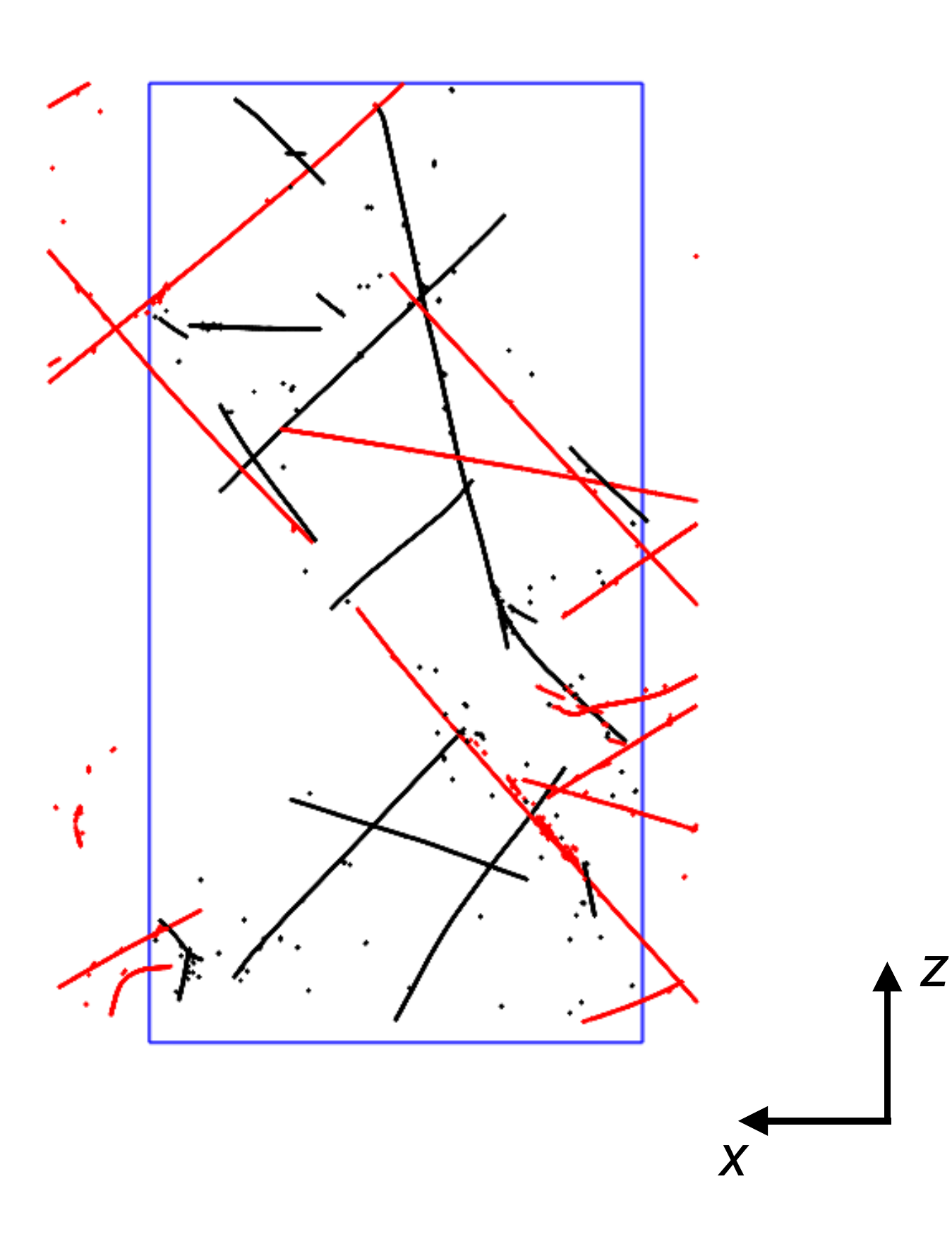}
\caption{The reconstructed output using the PandoraCosmic algorithm chain in 3D, the \textit{x-y} plane and the \textit{x-z} plane for a simulated event in ProtoDUNE-SP. For illustrative purposes, only hits appearing in the beam-side central drift volumes in ProtoDUNE-SP have been reconstructed. Particles in red are deemed to be out of time, as they appear outside the physical boundary of the drift volumes because no offset in the drift position has been applied. Particles in black are those deemed to be in time. Out-of-time particles are tagged as cosmic-ray muons.}
\label{fig:intime}
\end{figure*}

The reconstructed cosmic-ray hierarchies are evaluated, and any hierarchies that represent clear cosmic-ray muons are tagged as fully reconstructed, and are not considered in subsequent reconstruction steps. A clear cosmic-ray particle hierarchy is tagged if it satisfies at least one of the following criteria:

\begin{itemize}
\item Any hits in the reconstructed particle (placed assuming arrival at the beam trigger time) fall outside of the physical drift volume boundary, as illustrated by the red particles in Fig.~\ref{fig:intime}.
\item The reconstructed particle was stitched across the cathode or an APA plane and the difference between the reconstructed $t_0$ and the beam trigger time exceeds 6.2\,$\upmu$s, corresponding to a shift of 1\,cm in drift position in the stitching process.
\item The reconstructed particle crosses the top and bottom boundaries of the detector.
\item A track fitted to the reconstructed particle has a direction consistent with a downward-going cosmic ray with very little curvature. \footnote{Curvature here is defined as the average deviation of direction between pairs of hits in the track from the average direction of the track.}
\end{itemize}

A total of 62.4$\pm$0.04\% of cosmic rays are tagged using this method and identified as clear cosmic-rays muons. Those reconstructed particle hierarchies tagged as clear cosmic-ray muons are set aside to form one part of the consolidated reconstruction output. The hits that form these clear cosmic-ray muon hierarchies are not considered in the remaining steps of the reconstruction.

\subsubsection{Event Slicing}
\begin{figure*}[htb]
\centering
\includegraphics[width=0.48\textwidth,trim={3cm 0cm 3cm 3.2cm},clip]{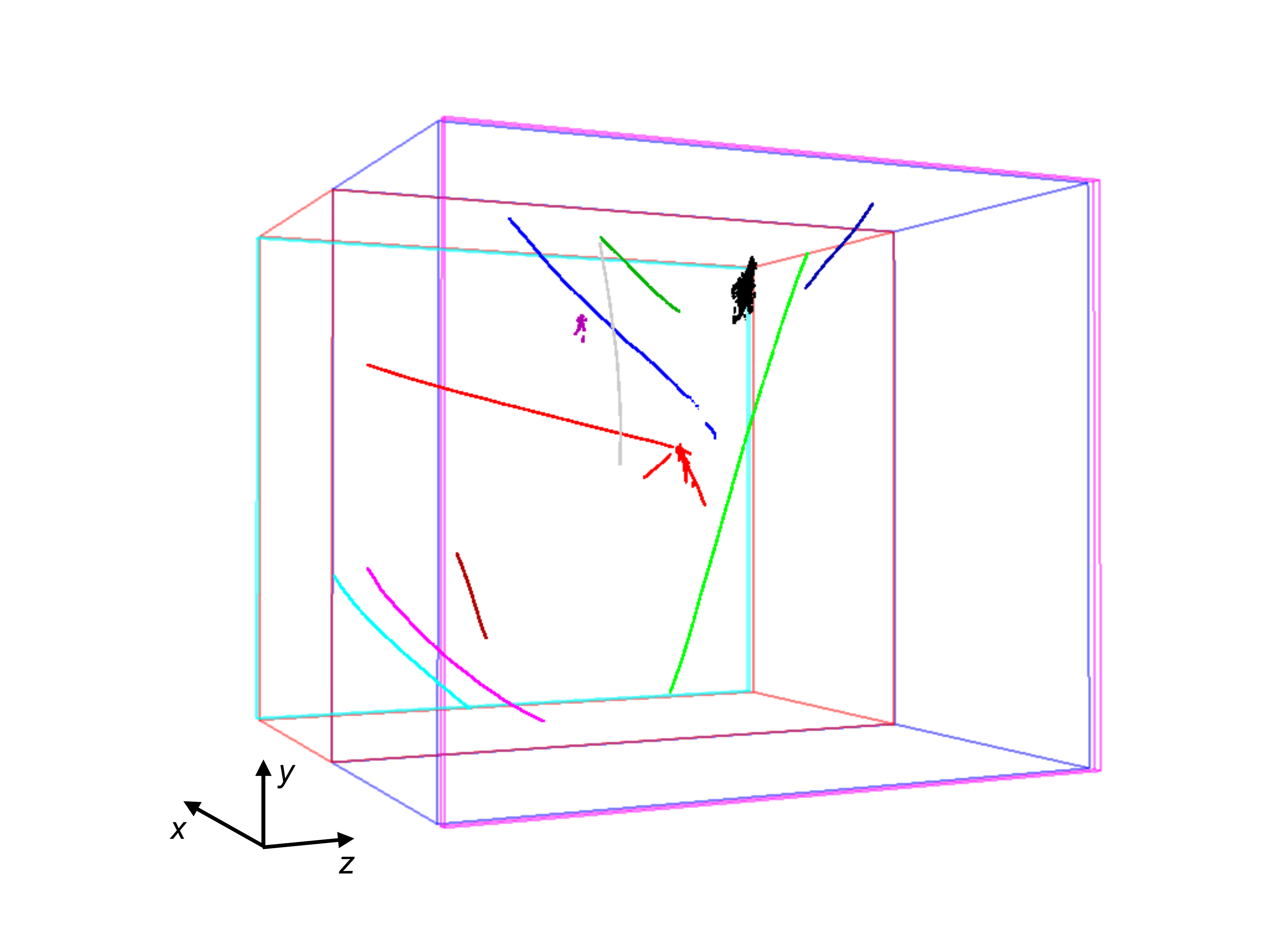}
\includegraphics[width=0.48\textwidth,trim={3cm 0 3cm 0},clip]{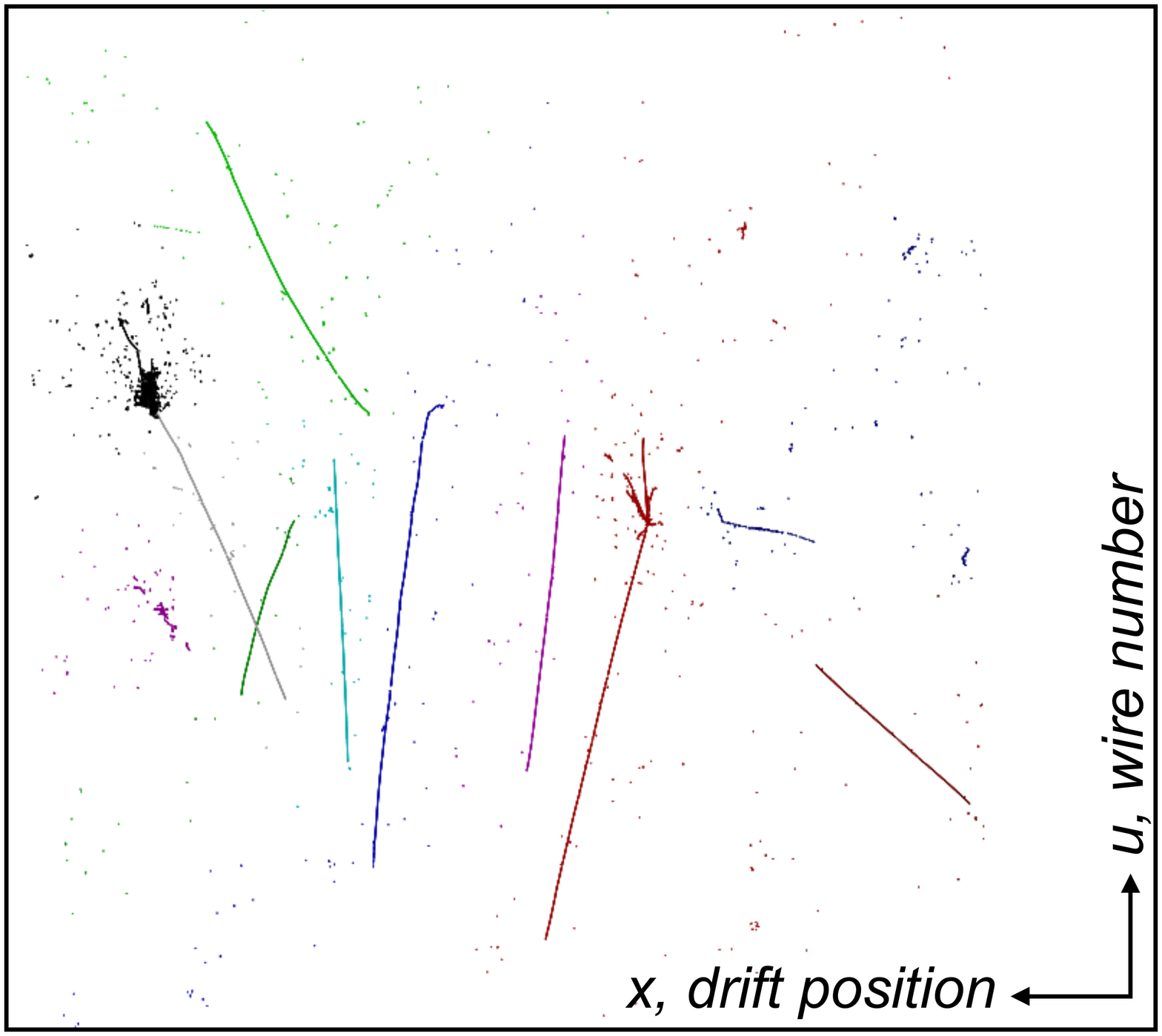}
\\
\includegraphics[width=0.48\textwidth,trim={3cm 0 3cm 0},clip]{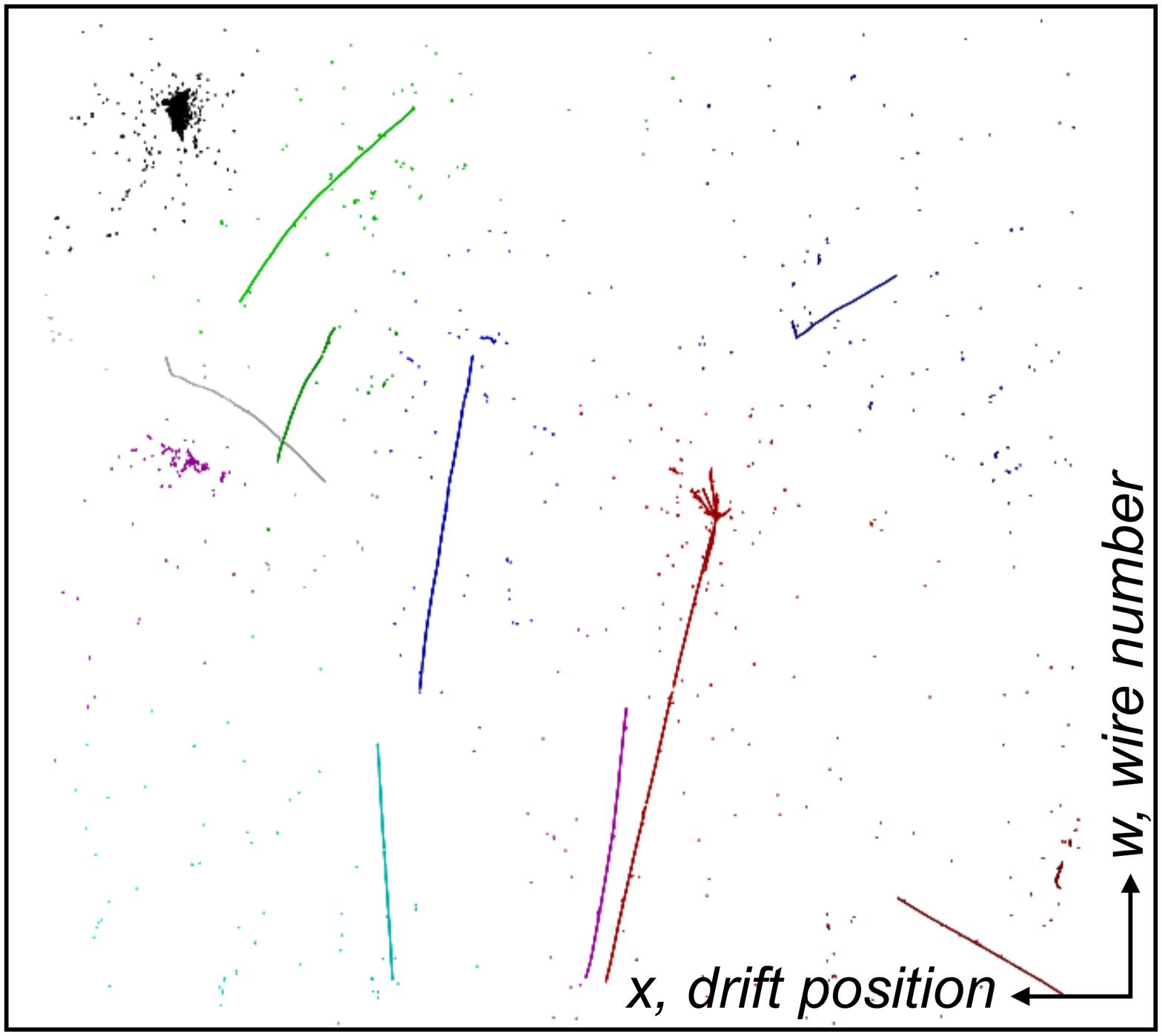}
\includegraphics[width=0.48\textwidth,trim={3cm 0 3cm 0},clip]{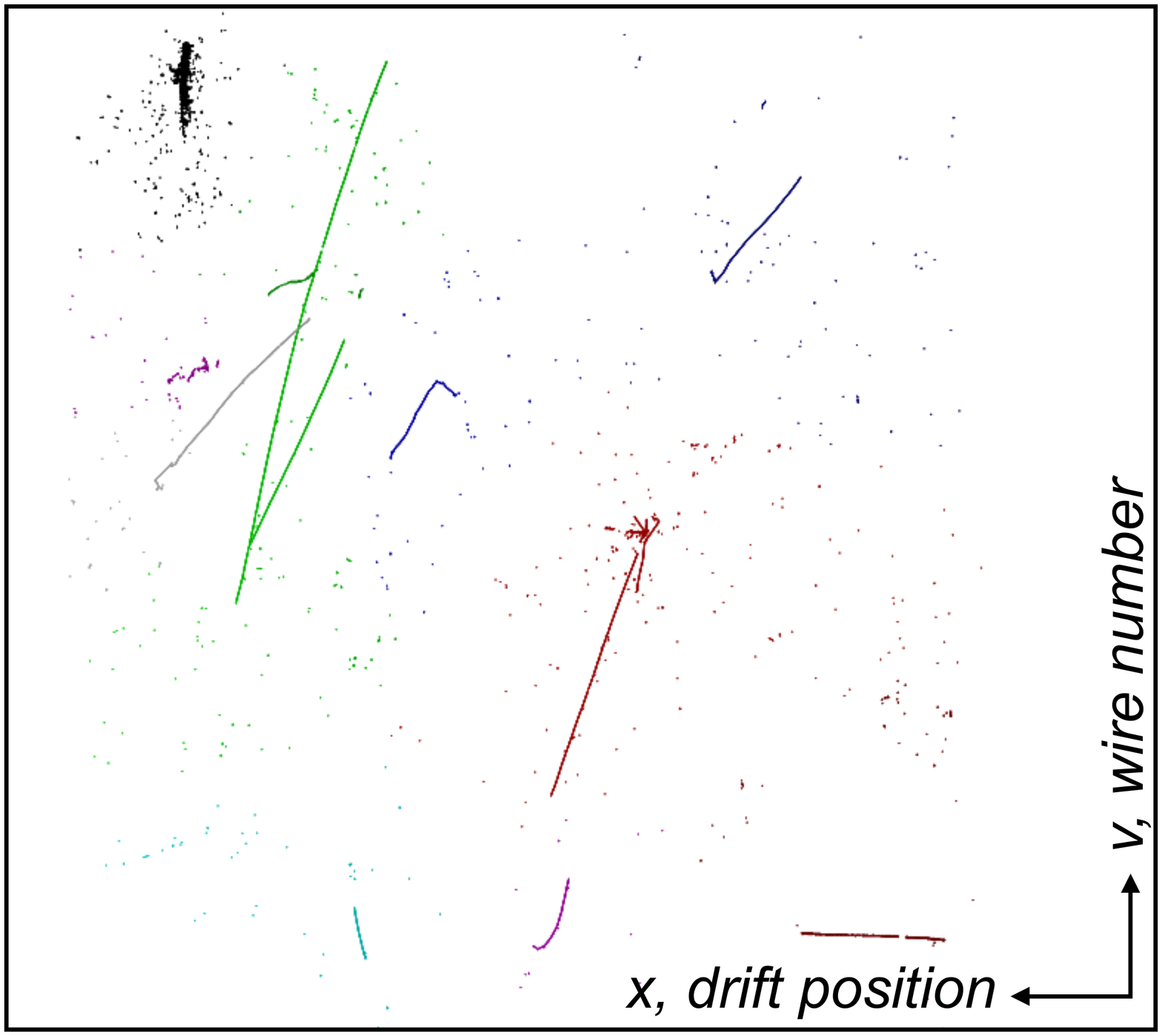}
\caption{The eleven ``slices'' created during the reconstruction of a simulated 3\,GeV/$c$ $\pi^+$ ProtoDUNE-SP interaction after the removal of the clear cosmic rays. Clockwise, from top left: 3D hits created by the `fast reconstruction'; and 2D hits in the $u$, $v$ and $w$ views. Each unique colour represents a distinct slice, and the reconstructed beam particle slice is shown in red.}
\label{fig:slicing}
\end{figure*}

The hits that do not form part of the clear cosmic-ray muon hierarchies are analysed further. The aim is to divide up the hits into \textit{slices}, where each slice contains hits from a single particle hierarchy. A subset of the PandoraTestBeam algorithms are used to perform a fast 3D reconstruction that allows the hits to be separated into groups arising from different primary particles. An example of the slicing procedure applied to a simulated interaction is shown in Fig.~\ref{fig:slicing} with the beam slice shown in red and a number of cosmic-ray slices. The output of the slicing algorithm is a list of hits produced for each reconstructed slice, and all hits that were input to the slicing algorithm must be assigned to a slice. In the remaining reconstruction stages the slices are processed separately.

\subsubsection{Slice Identification}

After the event slicing, different reconstruction hypotheses can be applied to each slice. The idea is that, for each individual slice, the hypothesis that produces the most appropriate reconstruction outcome can be selected. Each slice will have exactly one outcome selected, and the consolidated event will be built from: i) the clear cosmic-ray muon hierarchies and ii) one selected outcome for each slice. Each slice is reconstructed independently by the PandoraCosmic and PandoraTestBeam algorithm chains, providing two possible reconstruction outcomes.

The next step is to select one of the two different reconstruction outcomes for each slice and in ProtoDUNE-SP this decision is made using a BDT. The following features are calculated for the two different slice outcomes, and both sets are used as inputs to the BDT:

\begin{itemize}
\item The number of reconstructed particles in the slice.
\item The distance between the point at which the test beam is expected to enter the detector and the closest 3D hit.
\item The vertical distance between the top face of the detector and the closest reconstructed 3D hit.
\item The eigenvalues of the covariance matrix from a principal component analysis of the reconstructed 3D hits.
\item The opening angle between the principal axis of the reconstructed 3D hits and the expected direction of the test beam.
\end{itemize}

These features are motivated by the fact that the entrance position and direction of the triggered test-beam particles are well understood. Cosmic-ray muons typically enter through the top face of the detector and produce simple track-like topologies in the detector in contrast to the typically more complex test-beam particle interactions.

A threshold is applied to the output score from the BDT and those slices with scores exceeding the threshold are classified as test-beam particles, while all remaining slices are classified as cosmic-ray muons. The PandoraTestBeam reconstruction outcome is selected for all slices classified as test-beam particles, while the PandoraCosmic reconstruction is chosen for all slices classified as cosmic-ray muons. The slice identification results in reconstructed particles hierarchies identified as cosmic-ray muons or test-beam particles, which form the output of the consolidated reconstruction along with the clear cosmic-ray muons.

\begin{figure*}[htb]
\centering
\includegraphics[width=0.95\textwidth,trim={0cm 5cm 0cm 1cm},clip]{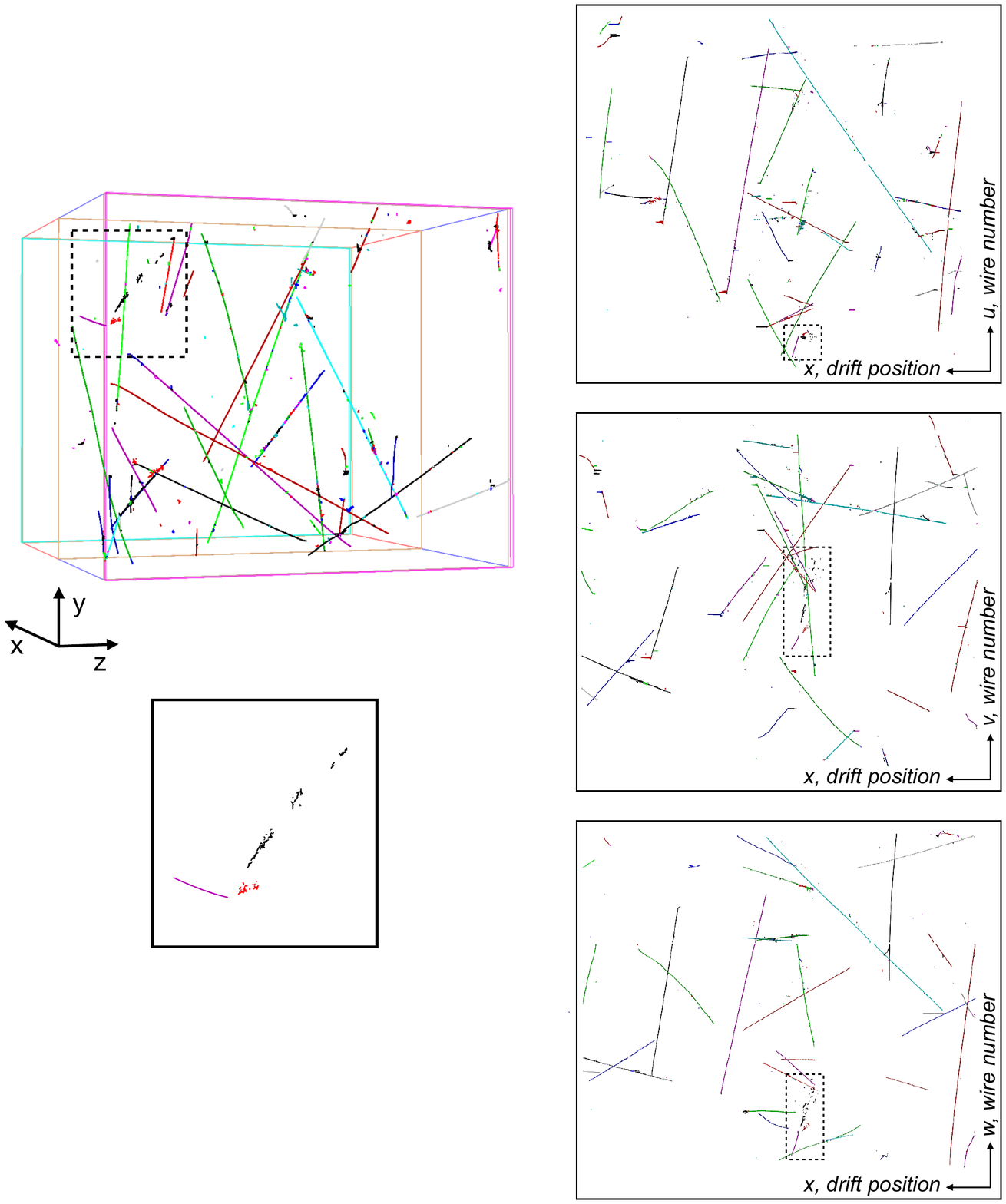}
\caption{The reconstruction output for a candidate 1\,GeV/$c$ $\pi^{+}$ charge exchange event from ProtoDUNE-SP data run 5387. The left image shows the 3D reconstruction output, highlighting the reconstructed particle hierarchy identified as the test-beam particle interaction: the reconstructed beam $\pi^+$ in purple comes from the left before interacting to produce two visible reconstructed $\pi^0$ decay photons in red and black. The figures on the right show, from top to bottom, the $u$, $v$ and $w$ view hits respectively for the fully reconstructed event with the beam particle interaction highlighted by the dashed black box.}
\label{fig:examplemcreco}
\end{figure*}

Figure~\ref{fig:examplemcreco} shows the reconstruction output for a candidate 1\,GeV/$c$ $\pi^{+}$ charge exchange event in ProtoDUNE-SP data, where the test-beam particle has been correctly distinguished from the cosmic-ray muon background. The zoomed view shows that the parent $\pi^{+}$ beam particle has been identified (purple, moving from left to right) and correctly placed at the top of the reconstructed hierarchy, and two $\pi^0$ decay photons emanating from the primary interaction vertex have been reconstructed (black and red) and added to the reconstructed hierarchy as child particles. Alongside the 3D reconstructed output, this figure also shows the 2D hits that form the reconstructed particles in the $u$, $v$ and $w$ views. 

\section{Simulated and Experimental Data}
\label{sec:datadescription}
Each event in simulation and data corresponds to one 3\,ms readout window of the detector, where the readout was initiated by the triggered test-beam particle. In addition, a number of background particles of both cosmic-ray and test-beam origin also traverse the detector, as shown in Fig.~\ref{fig:eventdecomp}. Unless otherwise specified, the simulated events include a data-driven simulation of the space charge effect.

The test-beam particle generation uses a GEANT4 simulation of the beamline~\cite{PhysRevAccelBeams.20.111001,PhysRevAccelBeams.22.061003}. The triggered test-beam particle is placed into the event with $t_0=t_\textrm{trigger}=0$ and other beam interactions are overlaid at random times, assuming a uniform distribution, to give background beam interactions spanning the entire 3\,ms detector readout window. Cosmic rays are simulated using CORSIKA \texttt{v7.4}~\cite{corsika} and are generated over a 6\,ms time range (centred on the trigger time) in order to completely cover the entire 3\,ms detector readout window. The simulation of particle propagation and interaction in the ProtoDUNE-SP detector is also performed by GEANT4, and the detector response simulation was performed using LArSoft \texttt{v08\_27\_01}~\cite{Church:2013hea}. The Pandora pattern recognition was performed using LArPandoraContent version \texttt{v03\_15\_02}, which in turn depends on version \texttt{v03\_03\_02} of the Pandora SDK.

The experimental test-beam data samples considered in this article were collected from August 2018 to December 2018. Due to time constraints, test-beam data were collected only in the positive polarity mode at five different particle momentum settings: 1, 2, 3, 6 and 7\,GeV/$c$. For this reason, only simulated interactions at these same five momentum settings are shown in this article.

Both data and simulation events go through signal processing and hit finding stages, as described in Ref.~\cite{proto_performance}. The events are input into Pandora after the reconstruction of hits from the signals identified on the detector readout wires. The average time taken to reconstruct a full ProtoDUNE-SP event with Pandora using the LArSoft framework is approximately 40\,s on an Intel Core Processor (Broadwell) 2.3\,GHz CPU while using an average of 2.8\,GB of memory.

\section{Cosmic Ray Reconstruction Performance}
\label{sec:cosmicrecoperformance}

The performance of the event reconstruction is first evaluated using the simulation and then compared to the experimental data. The method presented here to evaluate the performance of ProtoDUNE-SP event reconstruction for simulated interactions involves matching Monte-Carlo (MC) particles with reconstructed particles based on the number of \textit{shared hits}, which are those hits common to the reconstructed and true particles.
 
Selection criteria are applied to ensure that the MC particles are ``reconstructable'' and can be included in the performance metrics. The MC particles must produce at least 15 hits in the detector, with at least five hits in at least two of the three readout views. Furthermore, MC particle hits produced by non-primary neutrons, and photons produced by track-like primaries, that deposit energy a long way from the primary particle are not considered.

Matches are made by finding the match involving the largest number of shared hits between the reconstructed and MC particle. Once matched, the reconstructed and MC particles are declared unavailable for further matches. This process is then repeated for all remaining particles in the event. At this stage all reconstructed and MC particles have at most one match. Any remaining reconstructed particles that have no match are associated to the MC particle (that by definition must already have a single match) with which they share the most hits, irrespective of the number of matches the MC particle already has.

Once the reconstructed particles have been matched to the MC particles, the following metrics can be defined for each matched pair:
\begin{itemize}
\item \textbf{Efficiency}: The fraction of MC particles that are matched to at least one reconstructed particle. The Clopper-Pearson method~\cite{clopperPearson} is used to calculate the confidence interval on efficiency measurements presented in this article.
\item \textbf{Purity}: The fraction of hits in the reconstructed particle that are shared with the MC particle.
\item \textbf{Completeness}: The fraction of hits in the MC particle that are shared with the reconstructed particle.
\end{itemize}

When reporting the reconstruction efficiency, only matches with at least 50\% purity and 10\% completeness are considered to ensure that the reconstructed particle is predominantly associated with a single MC particle, and that the match is not of very low quality. These cuts are not applied when reporting the completeness and purity of matches.  

\subsection{Reconstruction Performance for Simulated Interactions}
\label{sec:crmetrics}
\begin{figure}[htb]
 \centering
 \includegraphics[width=0.48\textwidth,trim={0 0 1.3cm 0},clip]{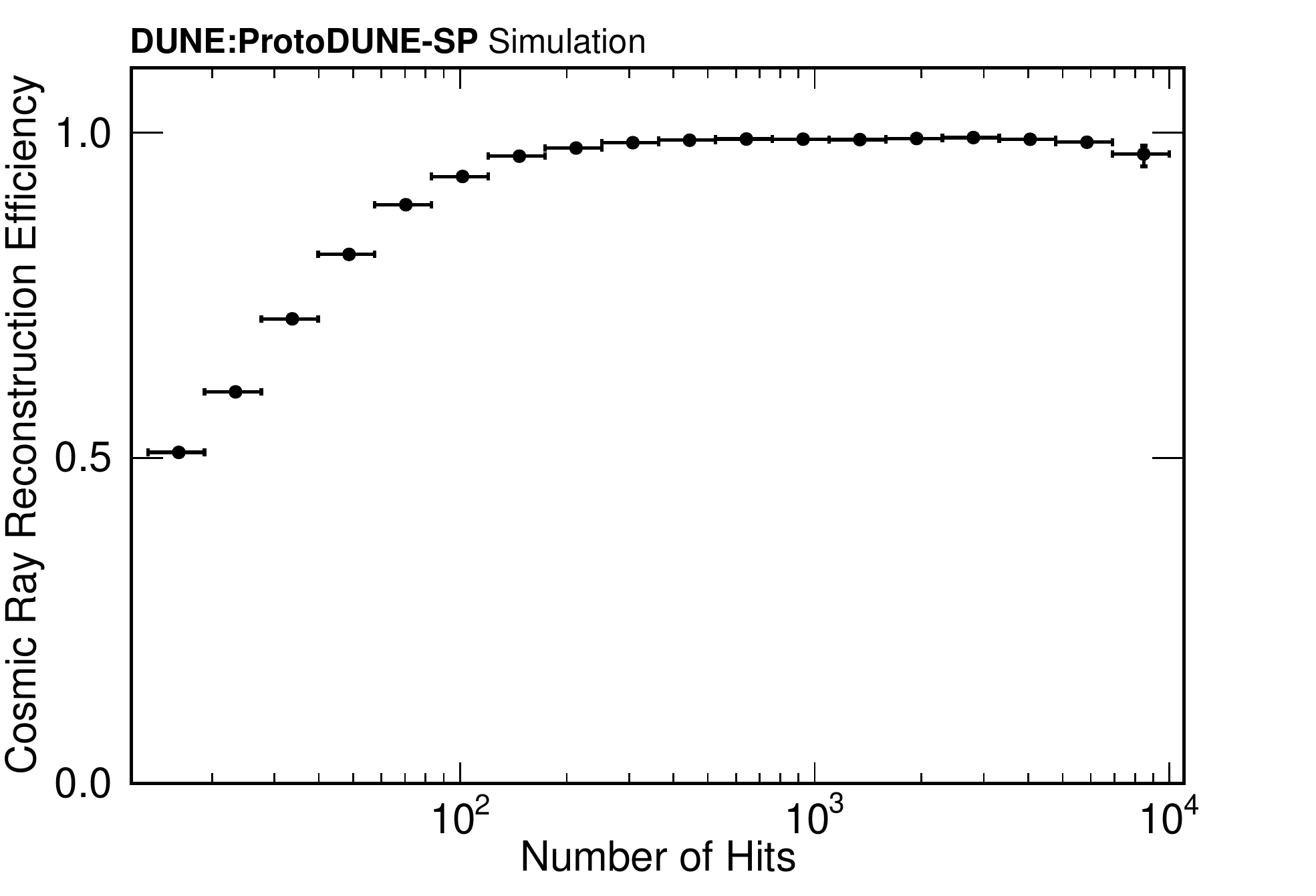}
 \includegraphics[width=0.48\textwidth,trim={0 0 1.3cm 0},clip]{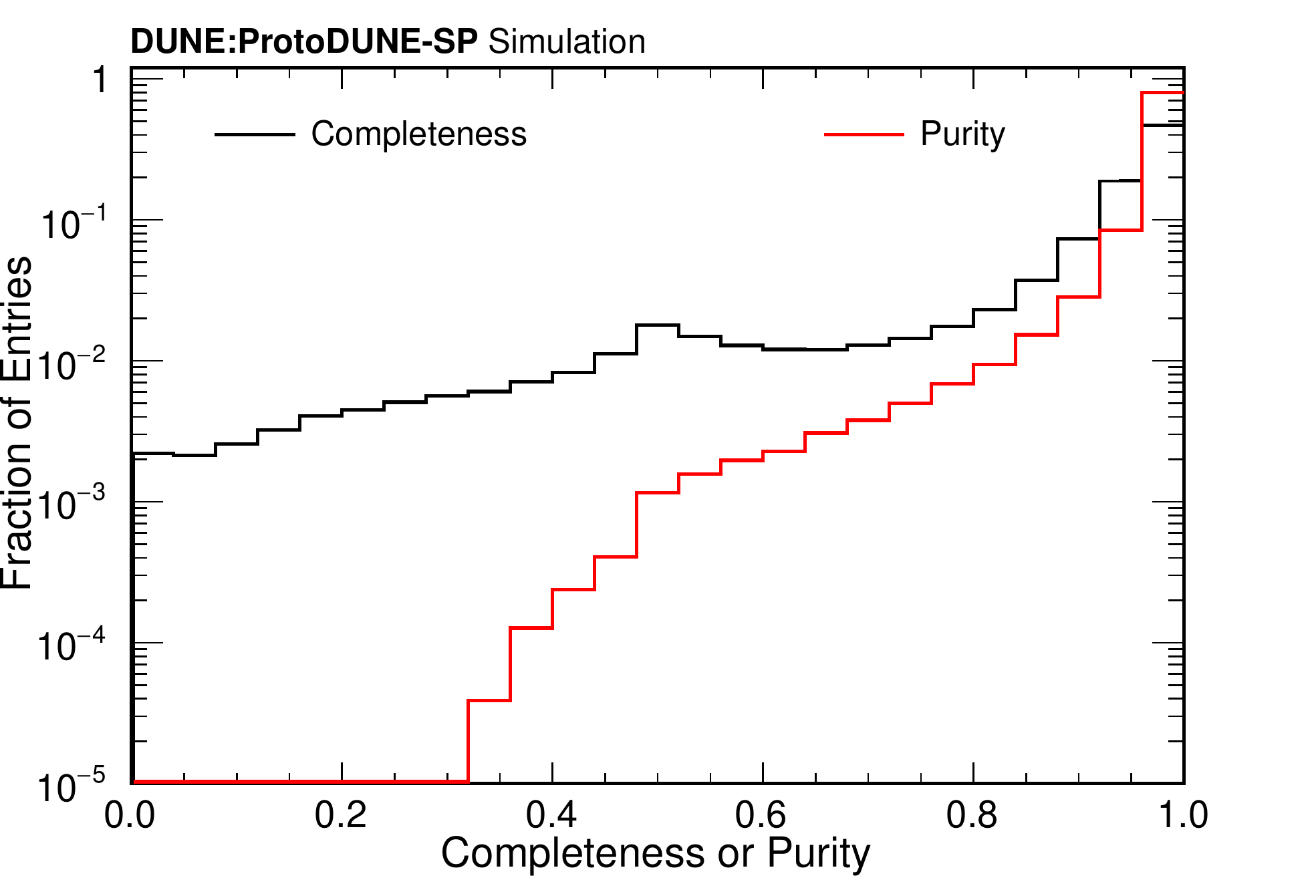}
 \caption{Left: the reconstruction efficiency for simulated cosmic-ray muons as a function of the true number of hits (summed over the three readout views) produced by the cosmic-ray muon. Right: the completeness and purity of the reconstructed cosmic-ray muons shown on a log scale.}
 \label{fig:crrecoeff}
\end{figure}

The left panel of Figure~\ref{fig:crrecoeff} shows the reconstruction efficiency for cosmic-ray muons as a function of the total number of true hits produced by the particle in the detector (including hits from delta-ray showers and Michel electrons). The overall integrated reconstruction efficiency for cosmic-ray muons is $95.73 \pm 0.03$\%. The reconstruction efficiency increases as a function of the number of hits, rising from 50\% for 15 hits up to 99\% for particles producing more than 400 hits. The reconstruction inefficiency for particles producing fewer hits is due to cosmic-ray muons being absorbed into larger neighbouring particles. This is more common for cosmic-ray muons producing a small number of hits, but it is also possible for long cosmic-ray muon track if the surrounding topology is sufficiently complex.

The completeness and purity of the reconstructed cosmic-ray muons are shown in right panel of Fig.~\ref{fig:crrecoeff}, both of which have very clear peaks at one. These figures show that 97.6\% of reconstructed cosmic-ray muons have a purity greater than 80\% and 81.9\% of reconstructed cosmic-ray muons have a completeness greater than 80\%. The tail on the low side of the completeness distribution
 is caused by the reconstruction splitting up a cosmic-ray muon track into two distinct particles. Approximately 8\% of the cosmic-ray muons are matched to two reconstructed particles, meaning that the reconstruction failed to reconstruct the particle as a single object. This can happen for a number of reasons, including failing to stitch the tracks at the drift volume boundaries, crossing cosmic-ray topologies and large delta-ray showers overlapping with with the muon tracks. The purity is typically close to 100\%, which indicates merging distinct cosmic-ray muons together is unlikely.

\begin{figure}[htb]
\centering
\includegraphics[width=0.60\textwidth,trim={0 0 1.3cm 0},clip]{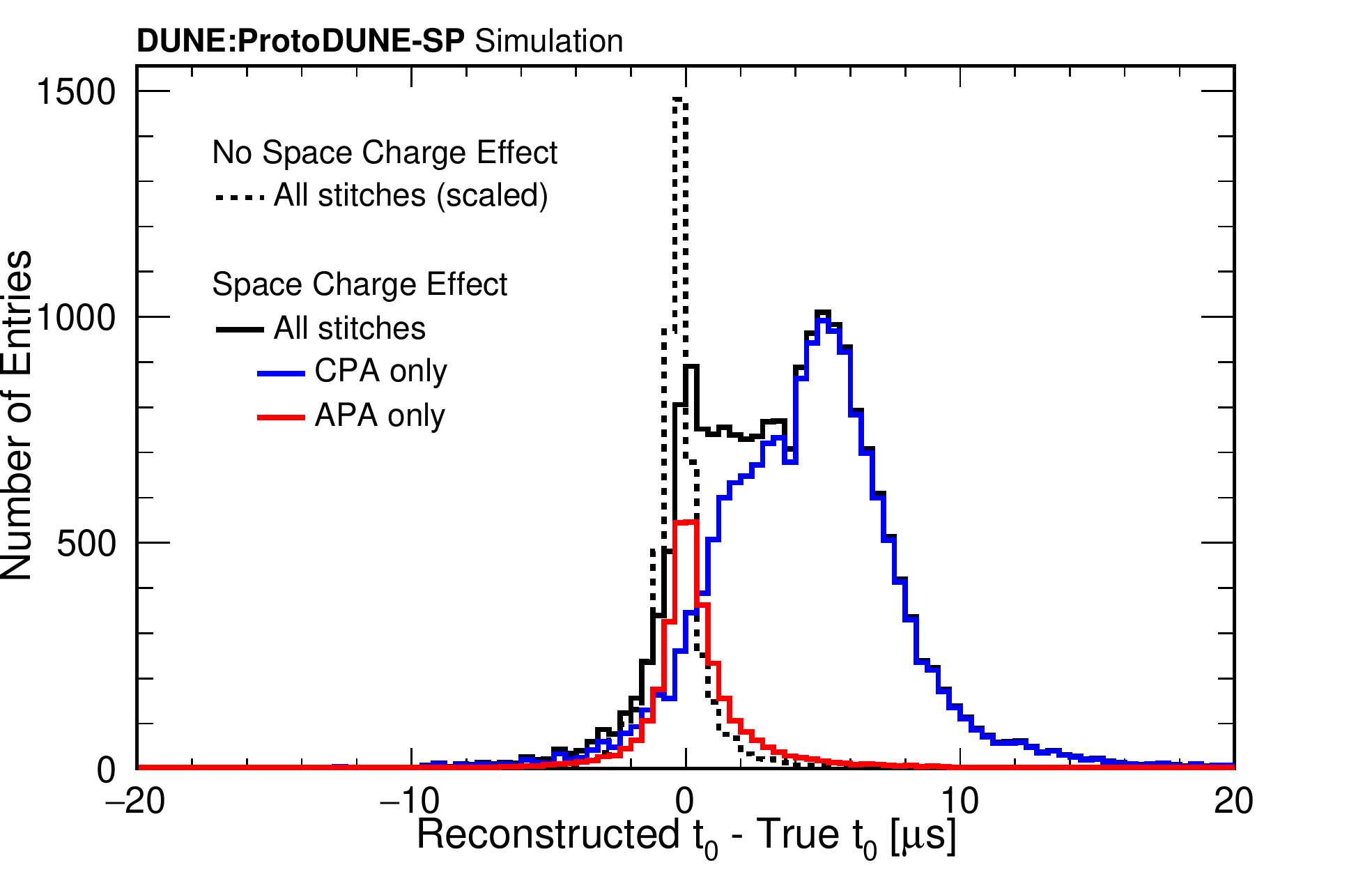}
\caption{The difference between the reconstructed and true $t_{0}$ for simulated cosmic-ray muons that have been stitched at either the CPA or APA with (solid black) or without (dashed black) space charge distortions. The black distribution is shown divided into the CPA- (blue) and APA-stitched (red) components. A time difference of 20\,$\upmu$s corresponds to shift of about 3\,cm in the drift direction.}
\label{fig:crt0res}
\end{figure}

It is possible to identify the time, $t_{0}$, that a cosmic-ray muon enters the LArTPC if the reconstructed particle was stitched between drift volumes by the process discussed in Sec.~\ref{sec:pandoracosmic}. The distribution of the $t_{0}$ residual, the difference between the reconstructed and true value of $t_{0}$, for stitched cosmic-ray muons is shown in Fig.~\ref{fig:crt0res}. The dashed black histogram shows the case where no space charge distortion was applied to the simulation and the distribution is centred on zero, as expected. Once space charge is included (the solid black distribution), a number of features become apparent when considering the cathode- (blue) and APA-stitched (red) components separately. The APA-stitched distribution remains centred on zero because the charge deposited close to the APA travels only a short distance and is unaffected by space charge distortions. Conversely, charges drifting from the cathode are maximally affected since they travel the entire drift distance, resulting in a distribution that is shifted by a few microseconds. Figure~\ref{fig:spacechargetrack} shows the effect of space charge on a reconstructed cathode-stitched cosmic-ray muon compared to the true trajectory. The bowing effect results in an overestimation of the shift in the drift direction, and hence the reconstructed $t_{0}$. Measurements of the SCE presented in Ref.~\cite{proto_performance} help to explain two further features of the distribution. The magnitude of the SCE varies across the LArTPC resulting in a broadening of the $t_{0}$ residual distribution. Finally, the asymmetric nature of the space charge distortions at the cathode causes a double-peak structure for cathode-stitched tracks, depending on whether the particle crossed the cathode from positive to negative $x$ or vice versa.

\begin{figure}
\centering
\includegraphics[width=0.48\textwidth]{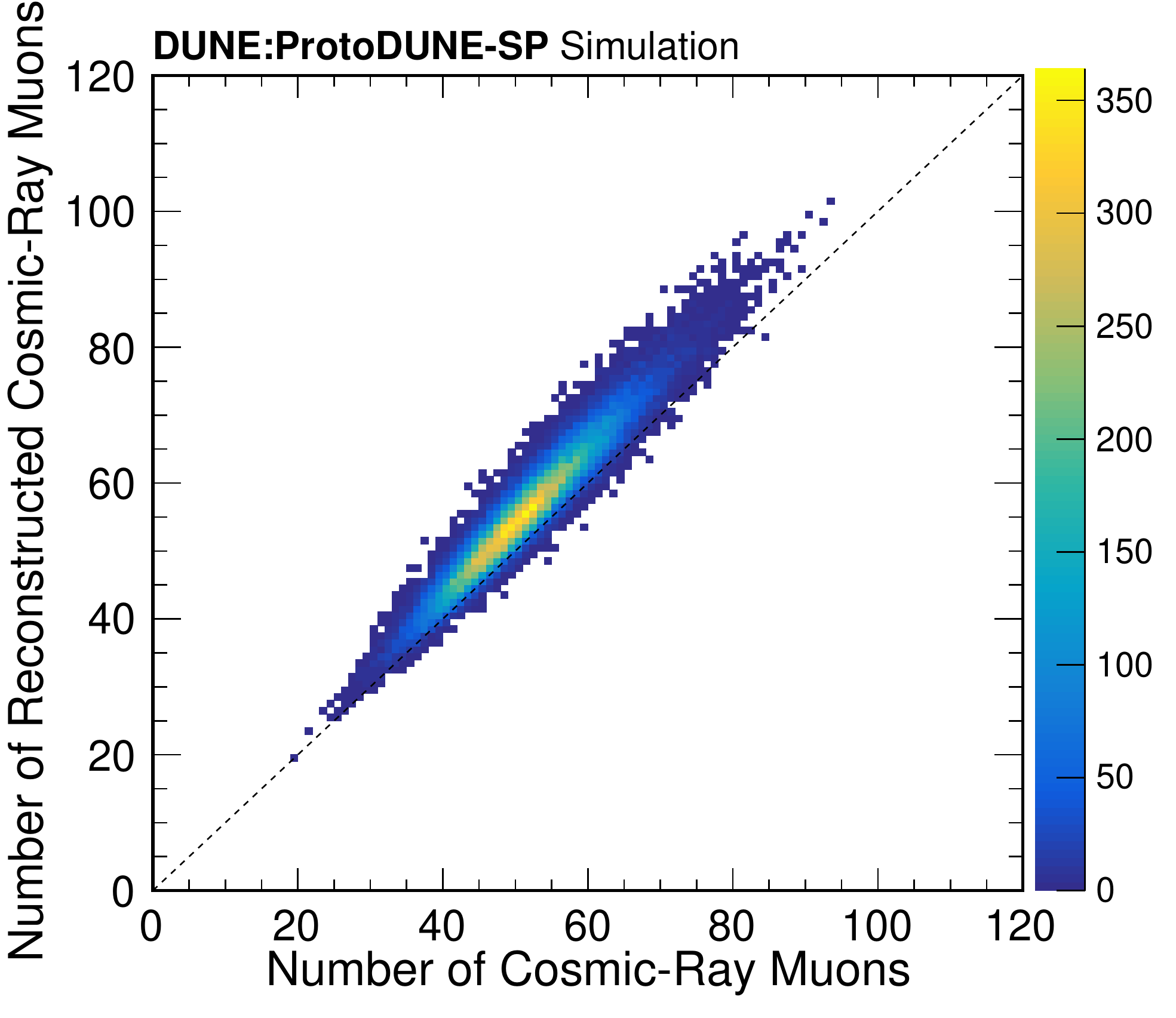}
\caption{The number of reconstructed cosmic-ray muons as a function of the number of true cosmic-ray muons on a per-event basis. The cosmic-ray muons were required to produce at least 100 hits in the detector.}
\label{fig:crnperevt}
\end{figure}

In order to give context to the topologies that the reconstruction is faced with at ProtoDUNE-SP, an estimate of the number of cosmic-ray muons passing through the detector per event in simulation has been made. The number of reconstructed cosmic-ray muons matched to distinct cosmic-ray muon MC particles, i.e. that deposit more than 100 hits in the detector, is shown as a function of the total number of distinct cosmic-ray muons on a per-event basis in Fig.~\ref{fig:crnperevt}. The distribution shows a strong linear correlation, but the gradient is approximately 1.08, corresponding to the aforementioned 8\% of cosmic rays that were reconstructed as two particles. However, it demonstrates that on average the cosmic-ray muons are well reconstructed. The mean number of distinct cosmic-ray muons per event is 52, while the mean number of matched reconstructed particles is 56, with negligible uncertainties.

\subsection{Reconstruction Performance for Cosmic-ray Data}
\begin{figure}[htb]
\centering
\includegraphics[width=0.48\textwidth,trim={0 0 1.3cm 0},clip]{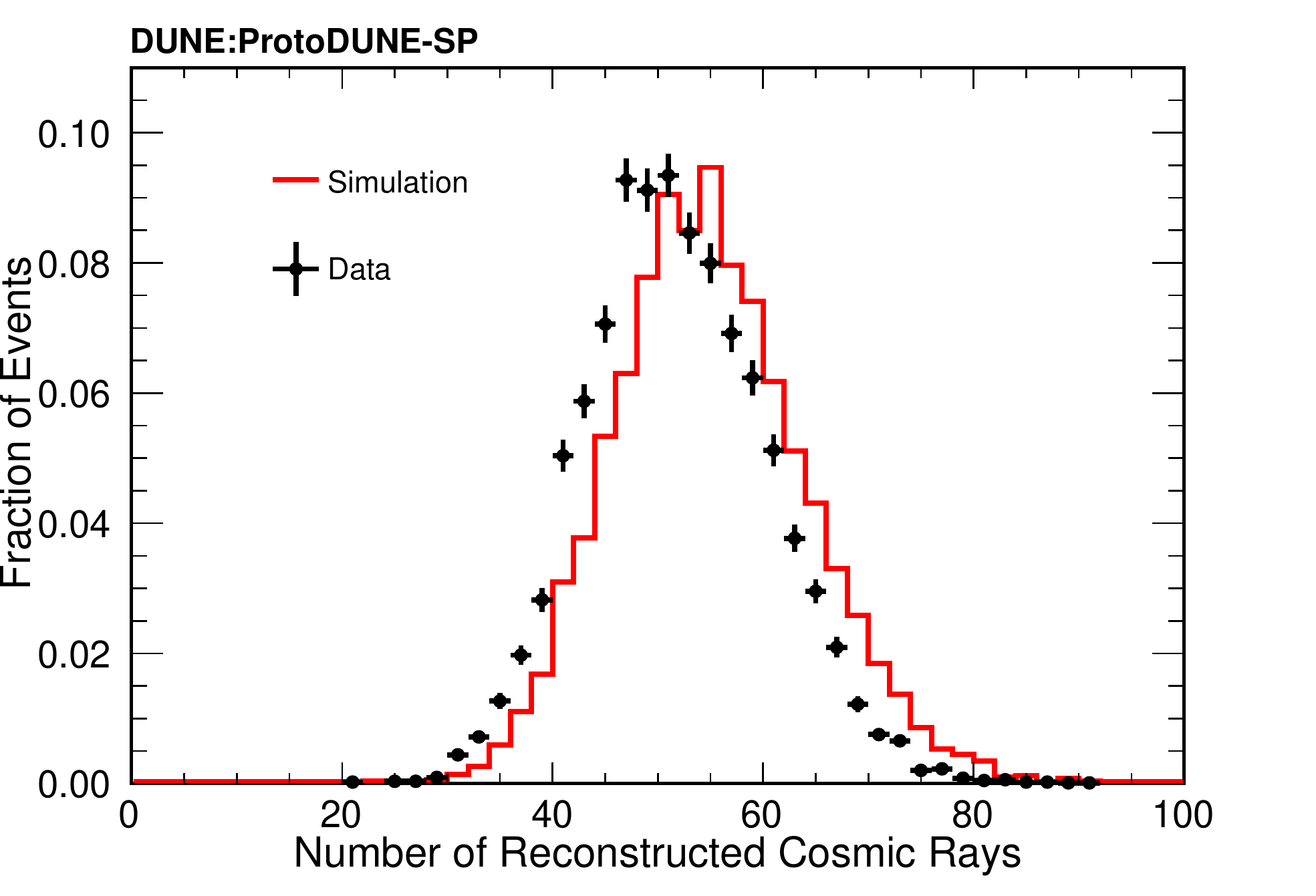}
\caption{The number of reconstructed distinct cosmic-ray muon particles per event for data (black) and simulation (red). The cosmic-ray muons were required to produce at least 100 hits in the detector.}
\label{fig:ncrdata}
\end{figure}

Reconstruction metrics for cosmic-ray muon data have also been evaluated. Figure~\ref{fig:ncrdata} shows the number of reconstructed particles tagged as distinct cosmic-ray muons per event in ProtoDUNE-SP. For a cosmic-ray muon to be tagged as distinct it must deposit at least 100 hits in the detector. This cut is applied in order to define a substantial, distinct signal in the detector. Furthermore, applying this cut yields a minimum reconstruction efficiency of \,90\%, based on the simulated efficiencies in Fig.~\ref{fig:crrecoeff}, which ensures this metric gives an accurate reflection of the true number of distinct cosmic-ray muons entering ProtoDUNE-SP. Approximately 5\% fewer cosmic-ray muons are reconstructed per event in data than simulation, with the data distribution peaking at $51.8\pm0.1$ and the simulated distribution peaking at $54.9\pm0.1$. This could be due to an overestimation of the cosmic ray flux in the simulation. Preliminary studies show that additional geometric selection criteria significantly improve the agreement in the mean number of reconstructed particles between data and simulation.

\begin{figure}[htb]
\centering
\includegraphics[width=0.7\textwidth]{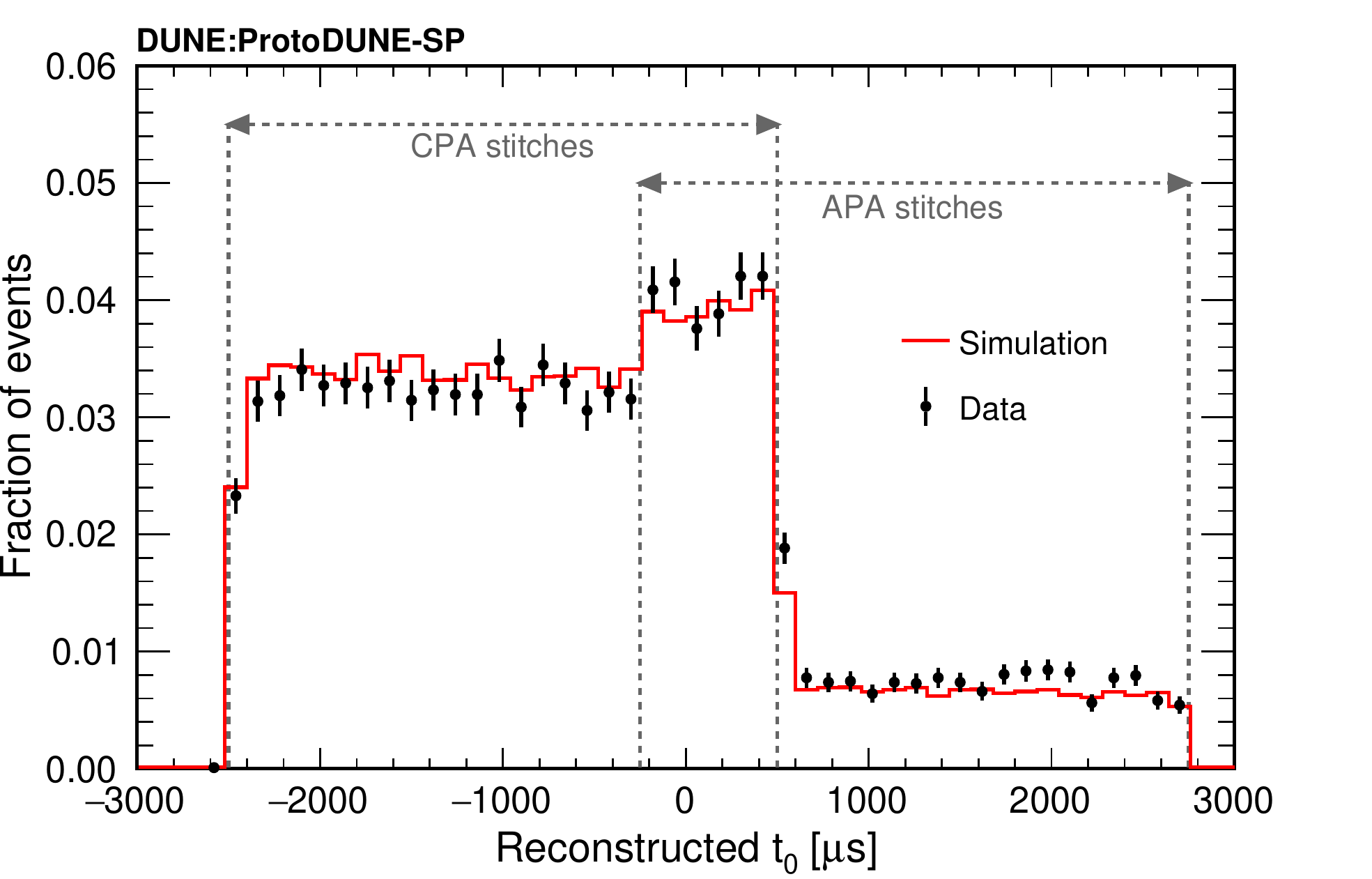}
\caption{The reconstructed $t_{0}$ distribution in ProtoDUNE-SP for cathode crossing and anode crossing cosmic-ray muons obtained from the Pandora stitching process in data and simulation. The distributions have been area normalised for comparison.}
\label{fig:recoT0data}
\end{figure}

The distribution of the reconstructed $t_{0}$ values for cathode crossing and anode crossing cosmic-ray muons is shown in Fig.~\ref{fig:recoT0data}. The range of this distribution can be predicted by considering the readout time window ($-$250\,$\upmu$s to 2750\,$\upmu$s) and the time for charge to drift from the cathode to the APAs (2250\,$\upmu$s). The cathode-crossing cosmic-ray muons have $t_0$ values in the range $-$2500\,$\upmu$s to 500\,$\upmu$s: the lower value is the start of the readout window minus the drift time, and the upper value is the end of the readout window minus the drift time. For APA-crossing cosmic rays, the $t_0$ values fall only within the readout window. Thus, the total distribution spans the range $-2500\,\upmu$s $< t_0 < 2750\,\upmu$s. Good agreement is seen between data and simulation and the distributions fall within the expected time window predicted above.

\section{Test-Beam Reconstruction Performance}
\label{sec:testbeamrecoperformance}

The reconstruction and identification of the triggered test-beam particle is a key part of the hadron cross-section analyses at ProtoDUNE-SP. This section evaluates the performance on simulation and experimental data.

\subsection{Reconstruction Performance for Simulated Interactions}
\label{sec:beam_mc}

\begin{figure}[htb]
    \centering
    \includegraphics[width=0.48\textwidth,trim={0 0 0.5cm 0},clip]{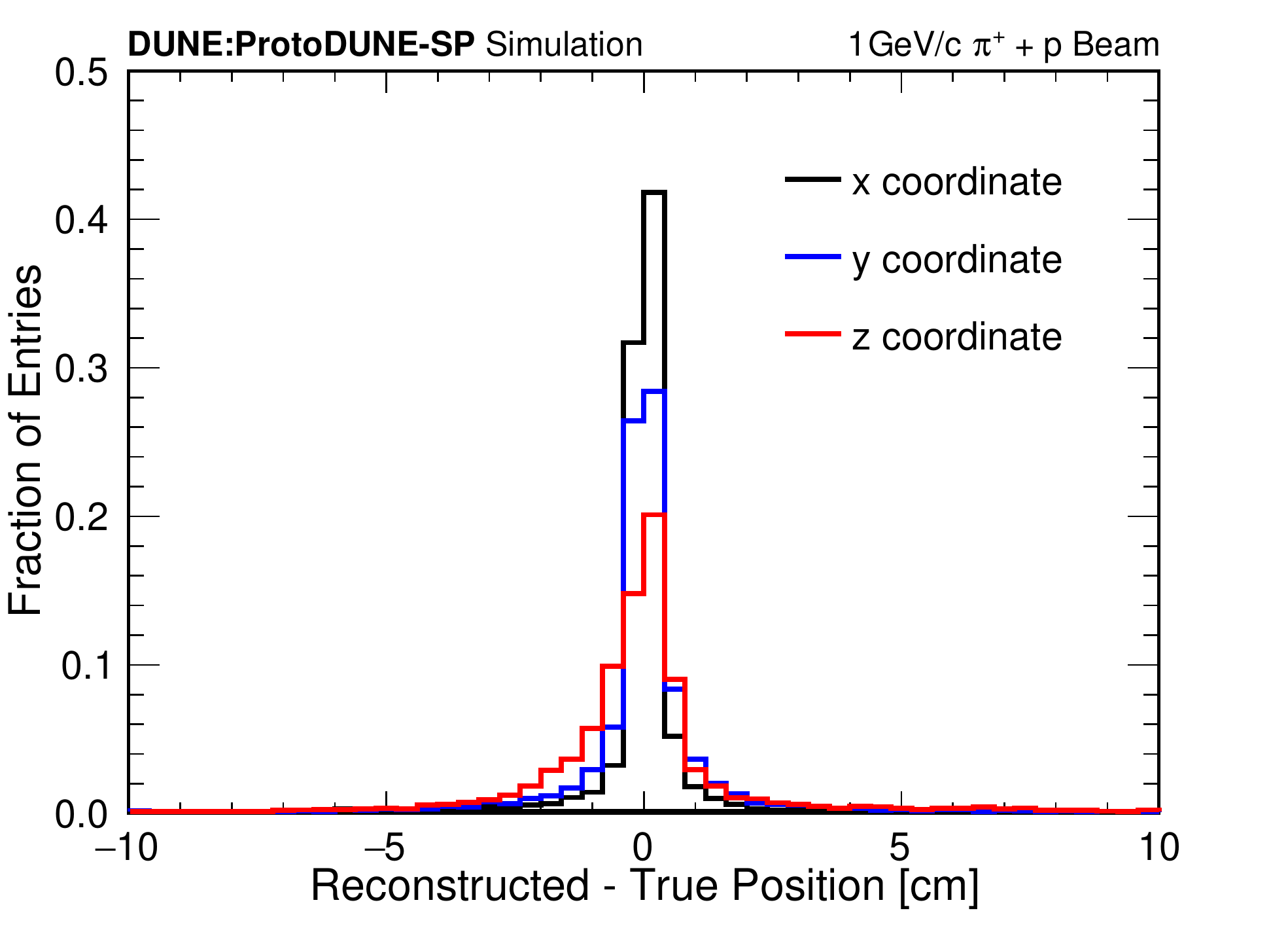}
    \includegraphics[width=0.48\textwidth,trim={0 0 0.5cm 0},clip]{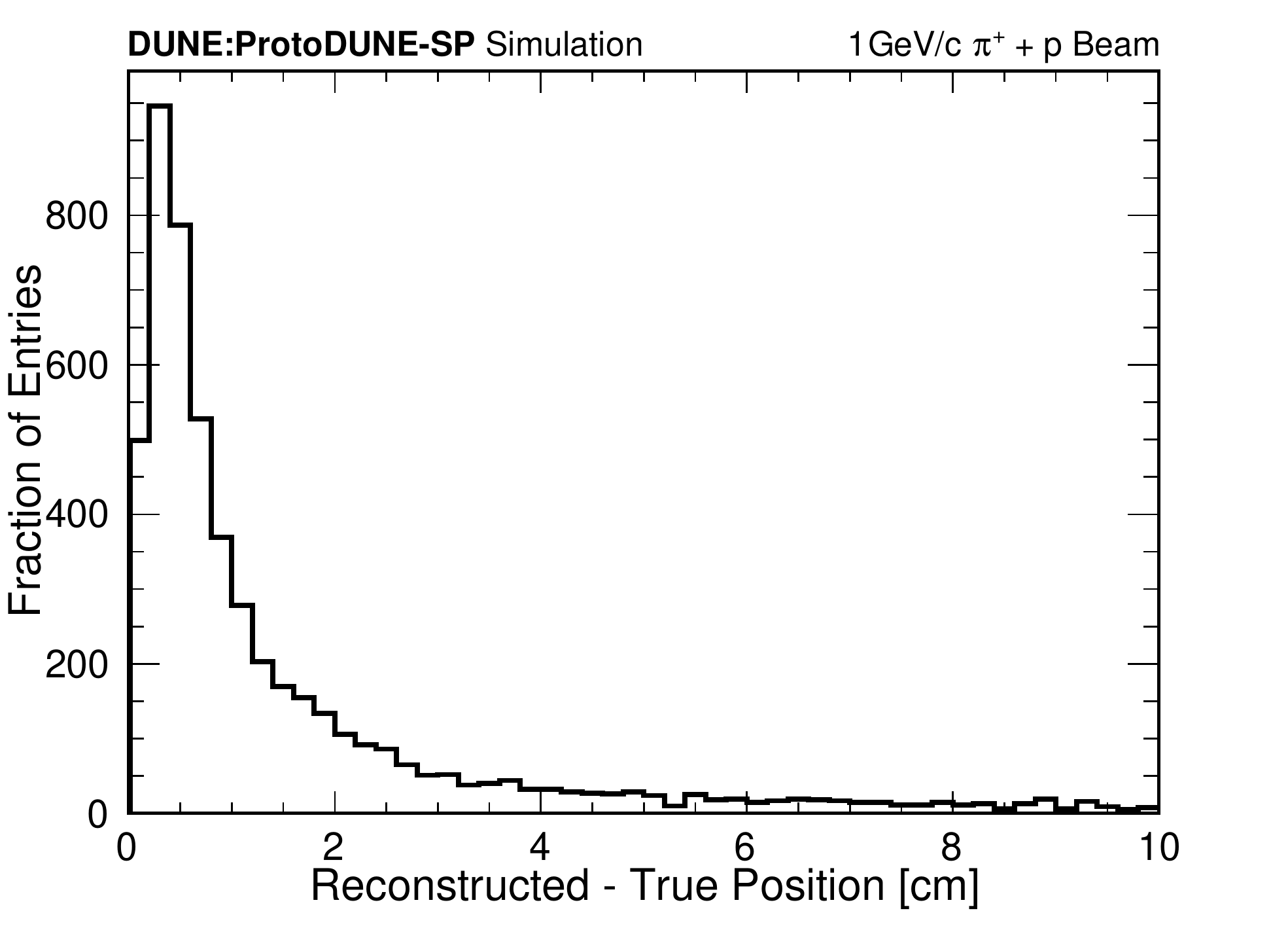}
    \caption{Left: the difference between the reconstructed and true end position of 1\,GeV/$c$ primary proton and charged pion test-beam particles shown for the $x$ (black), $y$ (blue) and $z$ (red) coordinates. Right: the three dimensional distance between the reconstructed and true end points.}
    \label{fig:end_point_res}
\end{figure}

The reconstruction of the triggered test-beam particle end point is of particular interest for cross-section analyses because it is critical to know where the particle either interacted or stopped~\cite{Gramellini:2018mjg,LArIAT:2021yix}. The differences between the reconstructed and true values for the end position coordinates of these particles are shown in Fig.~\ref{fig:end_point_res} for 1\,GeV protons and positively charged pions. The end point was corrected for SCE distortions using the procedure described in Ref.~\cite{proto_performance} and the resulting distributions are narrow and centred on zero, indicating good resolution and low bias. The right distribution shows the difference between the reconstructed and true positions in 3D, where $68\%$ of the beam particle end points are reconstructed within 2\,cm of the true value. 
 
 The efficiency to fully reconstruct triggered test-beam particles and to correctly identify them as of beam origin has been studied. In addition to the full simulation (including the triggered test-beam particle, beam-halo particles and cosmic rays), two additional simulated samples were used to understand the potential loss of efficiency due to background particles. The \textit{cosmics removed} sample has the cosmic rays removed from the event and hence consists only of the triggered test-beam particle and any beam-halo particles, and the \textit{cosmics and halo removed} sample further removes the beam-halo particles from the event, meaning only the triggered test-beam particle remains.

\begin{figure*}
\centering
  \includegraphics[width=0.48\textwidth,trim={0 0 0.5cm 0},clip]{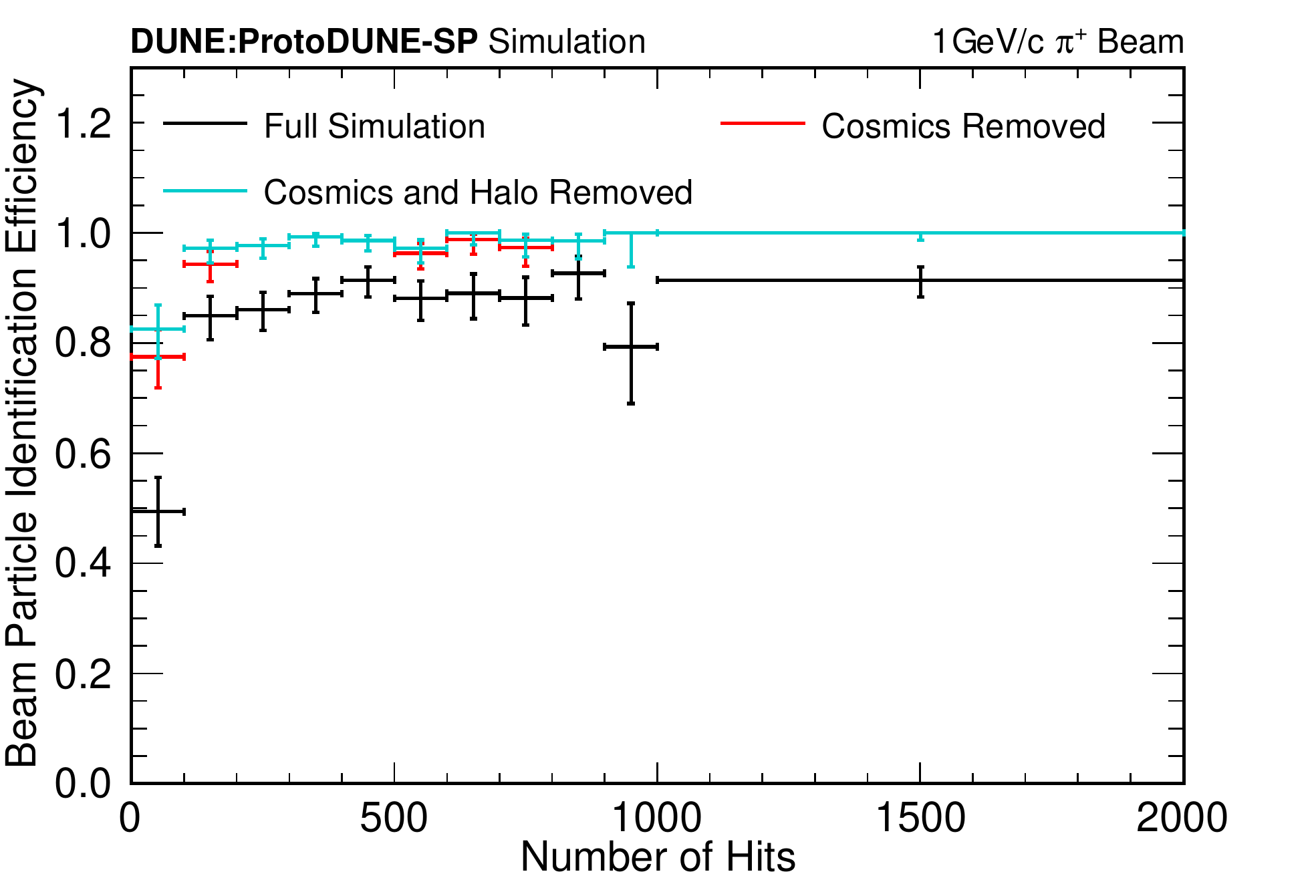}
  \includegraphics[width=0.48\textwidth,trim={0 0 0.5cm 0},clip]{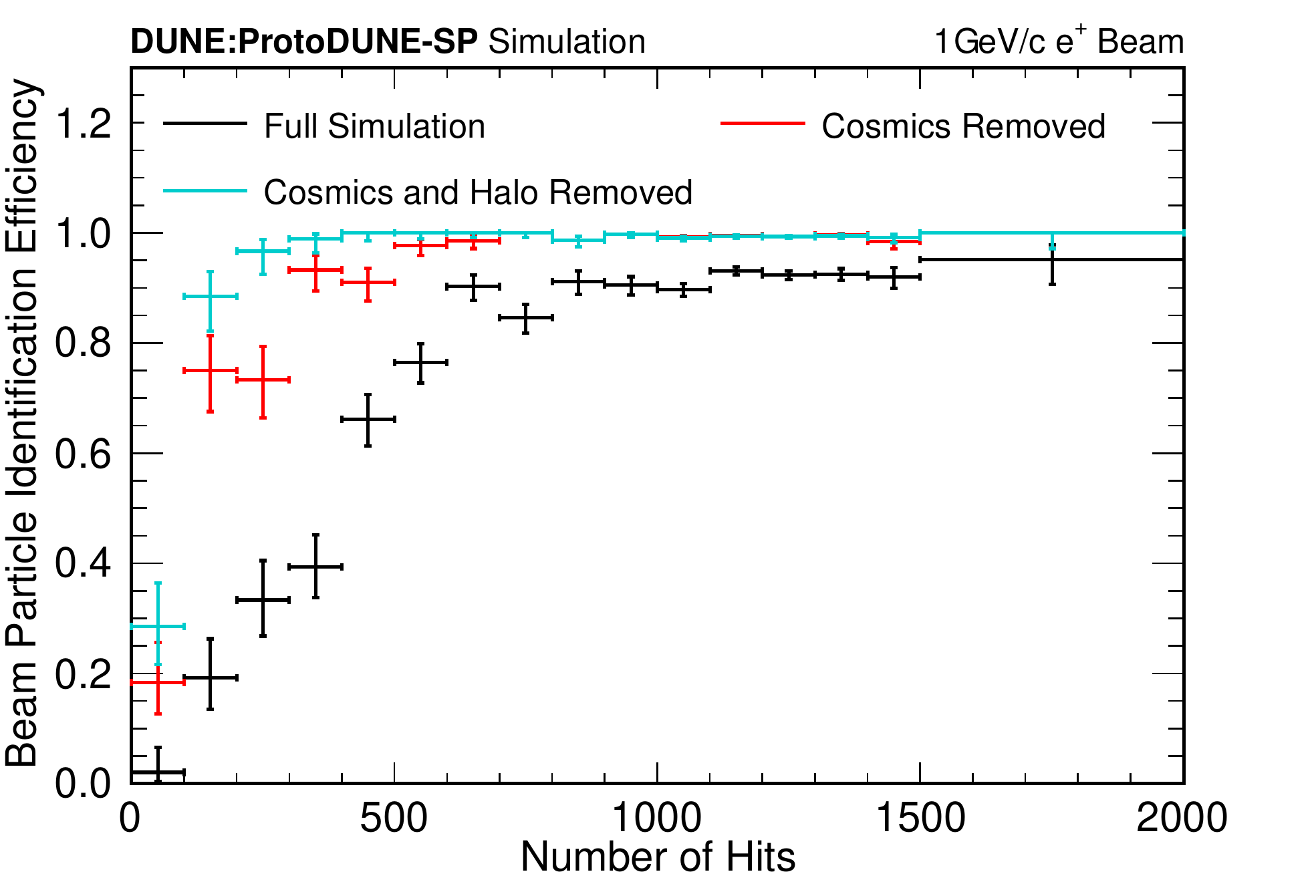} \\
  \includegraphics[width=0.48\textwidth,trim={0 0 0.5cm 0},clip]{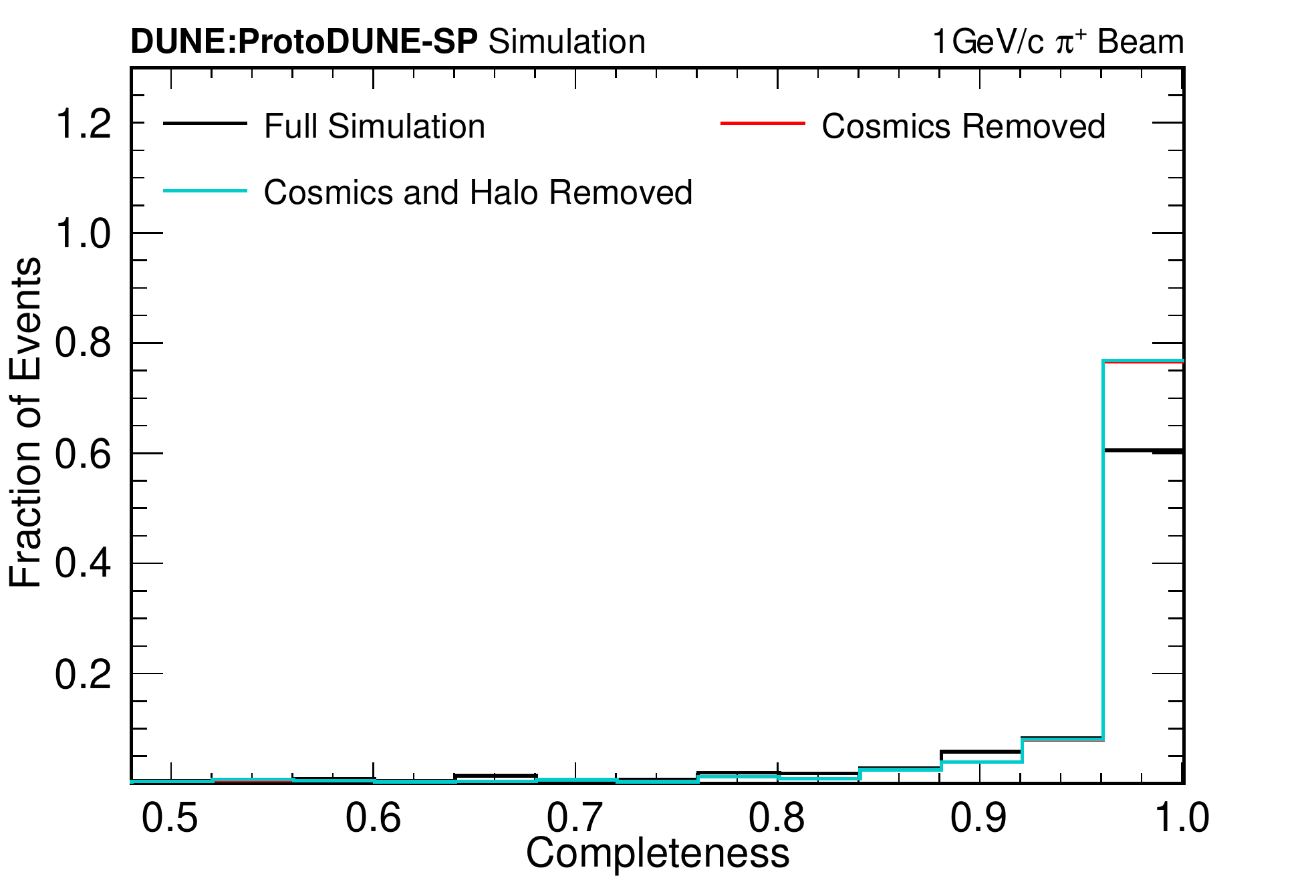}
  \includegraphics[width=0.48\textwidth,trim={0 0 0.5cm 0},clip]{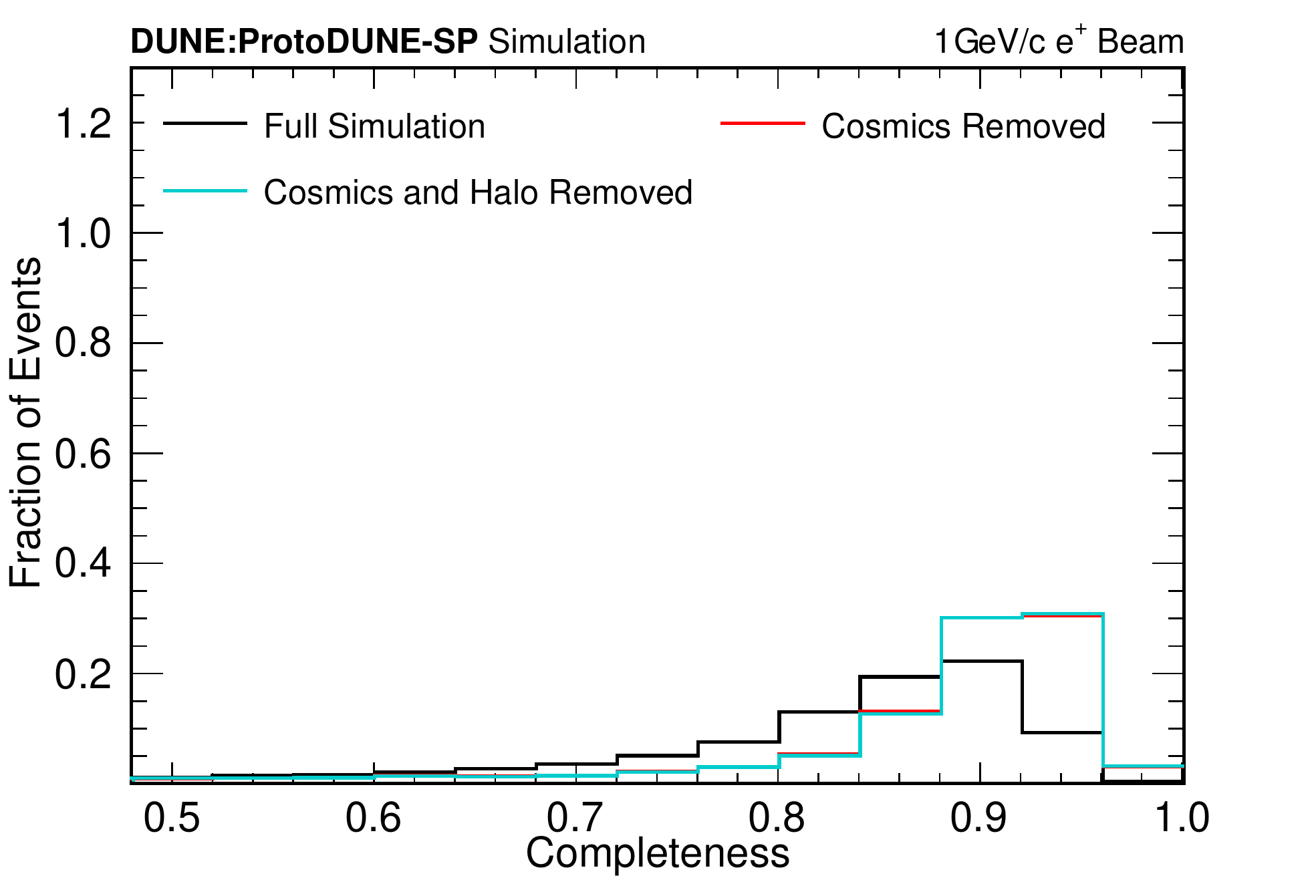}
  \includegraphics[width=0.48\textwidth,trim={0 0 0.5cm 0},clip]{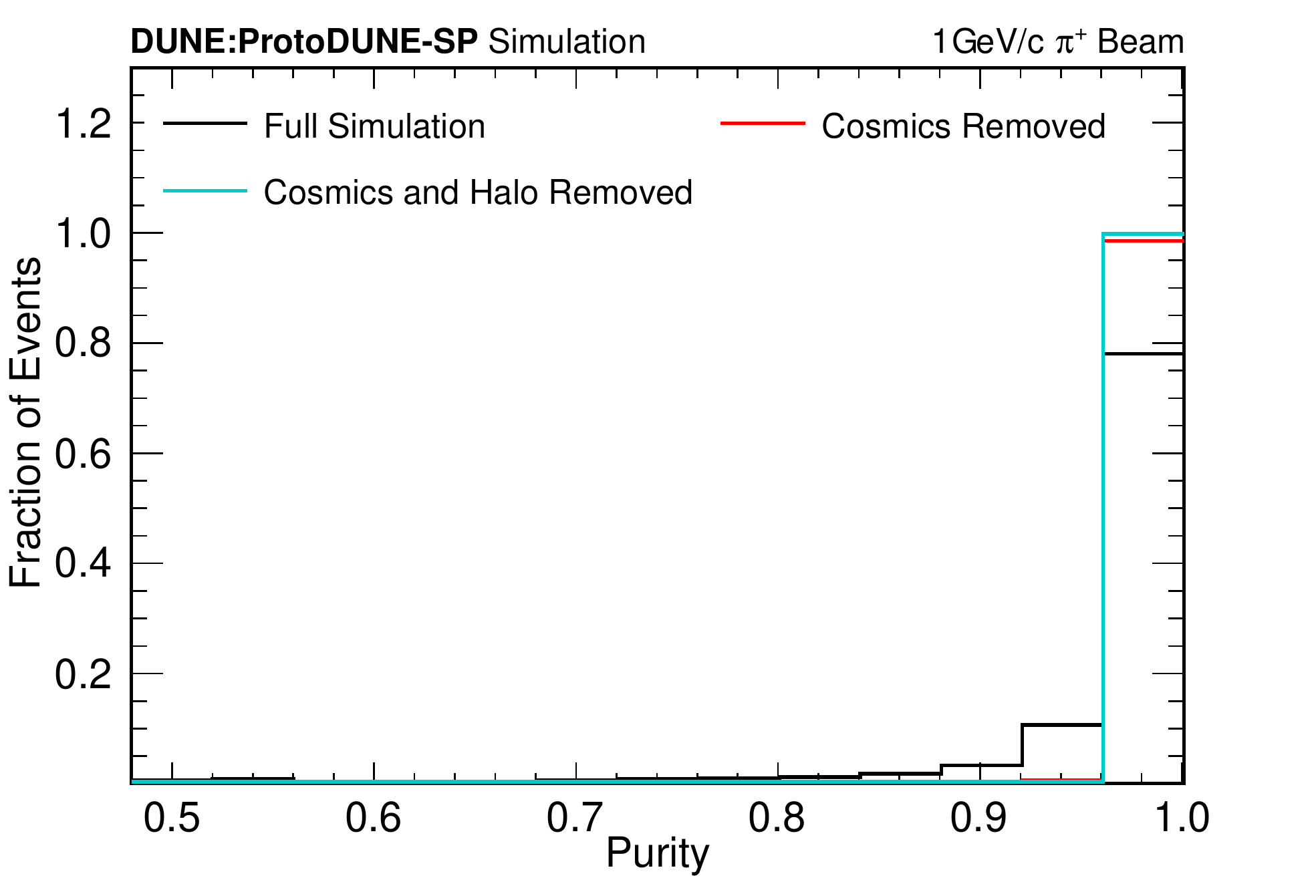}
  \includegraphics[width=0.48\textwidth,trim={0 0 0.5cm 0},clip]{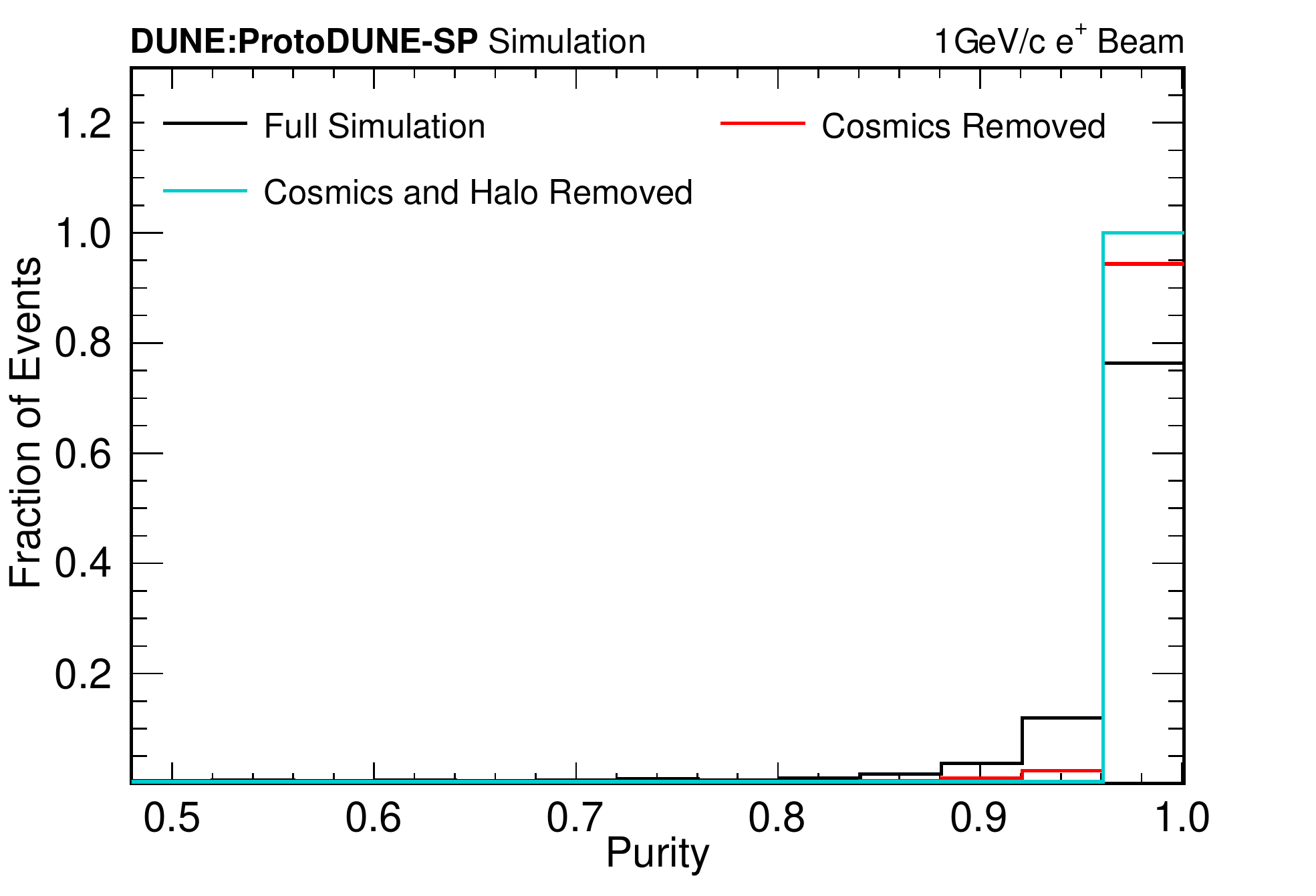}
\caption{Primary beam particle reconstruction and identification efficiency (top row) and the reconstructed particle completeness (middle row) and purity (bottom row) for 1\,GeV/$c$ $\pi^+$ (left column) and 1\,GeV/$c$ $e^+$ (right column) beam. The black distributions show the performance for the full ProtoDUNE-SP simulation, and the red and cyan curves show events with no cosmic rays, and no cosmic rays and no beam halo, respectively. In a number of places the red distribution is exactly covered by the cyan points.}
\label{fig:tbrecoeffbrkdwn}
\end{figure*}

 Figure~\ref{fig:tbrecoeffbrkdwn} shows six distributions that visualise the reconstruction performance for the standard simulation (black), the cosmics removed sample (red), and the cosmics and halo removed sample (cyan). Each column shows, from top to bottom: the triggered test-beam particle reconstruction and identification efficiency, meaning that the particle was well reconstructed and correctly identified as being the triggered test-beam particle; and the completeness and purity of the triggered test-beam particle and subsequent hierarchy. The left column is for 1\,GeV/$c$ $\pi^+$ interactions and the right column shows 1\,GeV/$c$ $e^+$ events.

\begin{figure}[htb]
\centering
\includegraphics[width=0.67\textwidth]{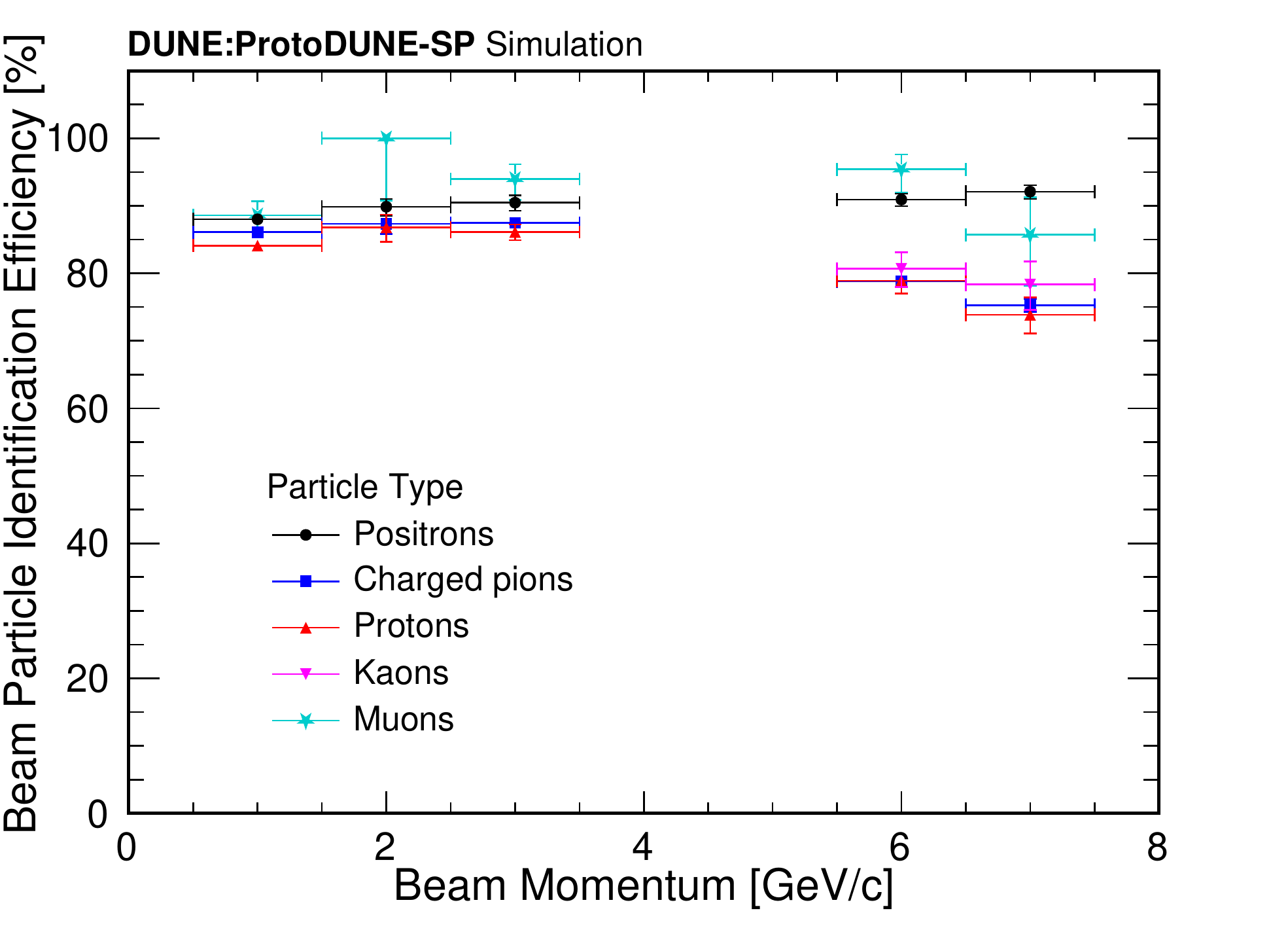}
\caption{The beam particle identification efficiency for the test-beam particle in ProtoDUNE-SP simulation for different beam momenta. No data were collected at 4\,GeV/$c$ or 5\,GeV/$c$, so no entries are shown for these momenta.}
\label{fig:beam_efficiency}
\end{figure}

The top figures show the reconstruction and identification efficiency for the triggered test-beam particles as a function of the number of 2D hits they produce in the detector (including hits produced by their interaction and decay products). The efficiencies for the full simulation both increase as a function of the number of hits and eventually plateau at $\sim$\,90\% for charged pions and $\sim$\,95\% for positrons. Removing cosmic-ray muons from the simulation significantly increases the efficiency over the whole range of the number of hits. At 1\,GeV/$c$ there are few beam halo particles, so only a small efficiency increase is seen after the sequential removal of the beam halo. As expected, the efficiency is approximately 100\% after the removal of all background particles, demonstrating that the performance on the full simulation is limited by the physics of the interactions and complex overlapping topologies. Similar behaviour is seen for the other beam particle types and the different momentum setting values. The average reconstruction and identification efficiency of the full simulation sample, for all particle types and beam momentum settings, is given in Table~\ref{tab:beamsimeff} and shown graphically in Fig.~\ref{fig:beam_efficiency}. There are more beam halo particles in the 6 and 7\,GeV/$c$ beam samples, which reduce the efficiency for charged pions and protons because they can be hidden within a beam halo positron shower. The efficiency remains high for high-energy positrons since they produce very large electromagnetic showers. 

\begin{table*}[htb]
\centering
\caption{The reconstruction and identification efficiency for the triggered test-beam particle in ProtoDUNE-SP simulation for positrons, charged pions, protons and charged kaons for different beam momenta. Charged kaons are negligible in number from 1 to 3\,GeV/$c$. The simulated events include the triggered test-beam particle, beam halo particles and numerous cosmic rays.}
\label{tab:beamsimeff} 
\begin{tabular}{cccccc}
\hline\noalign{\smallskip}
Momentum & \multicolumn{5}{c}{Reconstruction Efficiency $\left[\textrm{\%}\right]$} \\
$\left[\textrm{GeV/$c$}\right]$ & Positrons $\left(e^+\right)$& Pions $\left(\pi^+\right)$ & Protons $\left(p\right)$& Kaons $\left(K^+\right)$& Muons $\left(\mu^+\right)$\\
\noalign{\smallskip}\hline\noalign{\smallskip}
1 & 88.0$^{+0.2}_{-0.2}$ & 86.1$^{+0.6}_{-0.6}$ & 84.1$^{+0.6}_{-0.6}$ & -- & 88.6$^{+2.1}_{-2.5}$\\
2 & 89.9$^{+1.2}_{-1.3}$ & 87.3$^{+1.3}_{-1.5}$ & 86.8$^{+1.9}_{-2.1}$ & -- & 100.0$^{+0.0}_{-9.2}$\\
3 & 90.5$^{+1.1}_{-1.2}$ & 87.5$^{+0.5}_{-0.5}$ & 86.1$^{+1.1}_{-1.2}$ & -- & 94.0$^{+2.2}_{-3.1}$\\
6 & 90.9$^{+0.9}_{-1.0}$ & 78.8$^{+0.6}_{-0.6}$ & 78.9$^{+1.8}_{-1.9}$ & 80.7$^{+2.5}_{-2.7}$ & 95.4$^{+2.2}_{-3.5}$\\
7 & 92.1$^{+1.0}_{-1.1}$ & 75.3$^{+0.9}_{-1.0}$ & 73.8$^{+2.6}_{-2.8}$ & 78.3$^{+3.4}_{-3.8}$ & 85.7$^{+5.5}_{-7.6}$\\
\noalign{\smallskip}\hline
\end{tabular}
\end{table*}

The middle and bottom rows of Fig.~\ref{fig:tbrecoeffbrkdwn} show the completeness and purity of the reconstructed test-beam particle hierarchy, respectively. The four distributions are peaked at, or close to, one, indicating that the triggered test-beam particle is being reconstructed as a single, complete particle. Removing the cosmic-ray muons significantly improves both completeness and purity of the test-beam particle reconstruction. This improvement is expected as there are fewer hits to contaminate the reconstructed beam slice, or to incorrectly split the reconstructed beam particle in the slicing algorithm. Removing both cosmic-ray muons and beam halo particles enforces a purity of 100\% as all possible sources of contamination have been removed, and there is a small increase in the completeness. The effect of removing the cosmic rays is larger, which is to be expected as there are significantly more cosmic rays than beam halo particles, such that the slicing algorithm, a source of incomplete triggered test-beam particles, is less active for events where cosmic-ray muons have been removed irrespective of the presence of the second-order beam halo effect.

\subsection{Reconstruction Performance for Test-beam Data}
\label{sec:data}

In order to compare the triggered test-beam particle reconstruction and identification efficiency between data and MC, a less strict definition of efficiently reconstructed test-beam particles is used in comparison to the one described in Section~\ref{sec:beam_mc}. The following selection is used to obtain the sample of events that are highly likely to contain a beam particle, and hence form the denominator for the efficiency calculation.

\begin{itemize}
\item[]Data:
\begin{enumerate}
  \item The beam trigger is active.
  \item A single track is reconstructed in the beam monitors immediately upstream of ProtoDUNE-SP.
  \item The high-voltage applied to the cathode is stable at -180\,kV.
  \item The readout electronics on the beam-side of the detector are active.
  \item There are at least 10 3D hits in the region where the beam particle enters the detector. These 3D hits are produced as part of the disambiguation procedure described in Sec.~\ref{sec:patrec}, not within the Pandora software.
\end{enumerate}
\item[] Simulation:
\begin{enumerate}
  \item There is a triggered test-beam particle in the MC particle hierarchy
  \item There are at least 10 3D hits in the region where the beam particle enters the detector.
\end{enumerate}
\end{itemize}
The efficiency is then defined as the fraction of the selected events with a reconstructed beam particle hierarchy. 

There are some limitations in the ability of the beam instrumentation alone to give the unambiguous particle identification required to calculate the denominator of the reconstruction and identification efficiency. Pions and antimuons are not distinguished, and since triggered test-beam charged pions can decay in the beamline to antimuons, they are included as a joint sample. For the 6 and 7\,GeV/$c$ samples, positrons, charged pions and antimuons can not be distinguished, and are hence not included in the comparisons.\footnote{Physics analysis is possible as positrons are reconstructed as showers and can be easily selected. However, calculation of the absolute efficiency in data is not possible.} A summary of the momentum settings used for data and MC comparisons is as follows:
$\pi^{+}$/$\mu^+$ from 1 to 3\,GeV/$c$, $e^+$ from 1 to 3\,GeV/$c$, $p$ from 1 to 7\,GeV/$c$, and $K^+$ for 6 and 7\,GeV/$c$.

Two additional simulation samples were produced to investigate the effect of potential systematic uncertainties.
\begin{enumerate}
    \item The SCE-off sample does not have a simulation of the space charge effect. It gives an estimate of potential efficiency mismodelling due to differences in the SCE between data and MC. Since a sample with increased SCE was not available, the efficiency difference from using the SCE-off sample was used to produce a symmetric band around the standard simulation efficiency values.
    \item A sample with the beam halo component reduced by 15\%. This sample was motivated by Ref.~\cite{PhysRevAccelBeams.22.061003} that shows that the MC overestimates the beam trigger rate, and hence the event pileup. The difference between the efficiency measured from the standard simulation and this sample was taken as a symmetric systematic uncertainty centred on the standard simulation efficiency.
\end{enumerate}

Furthermore, to account for potential differences in the 3D hit finding between data and MC, the selection criterion requiring 10 3D hits in the region where the beam enters the detector was varied for the simulation to 8 and 12 and the change in efficiency was taken as a systematic uncertainty, giving the higher and lower limits on the systematic uncertainty band, respectively. This variation only has a significant effect for the 1\,GeV/$c$ sample because the low energy particles produce fewer hits than those at higher energies. The total systematic uncertainty is calculated as the quadrature sum of the aforementioned individual systematic uncertainties under the assumption of Gaussian uncertainties.

\begin{figure*}[htb]
\centering
\includegraphics[width=0.98\textwidth,trim={0 0 0.5cm 0},clip]{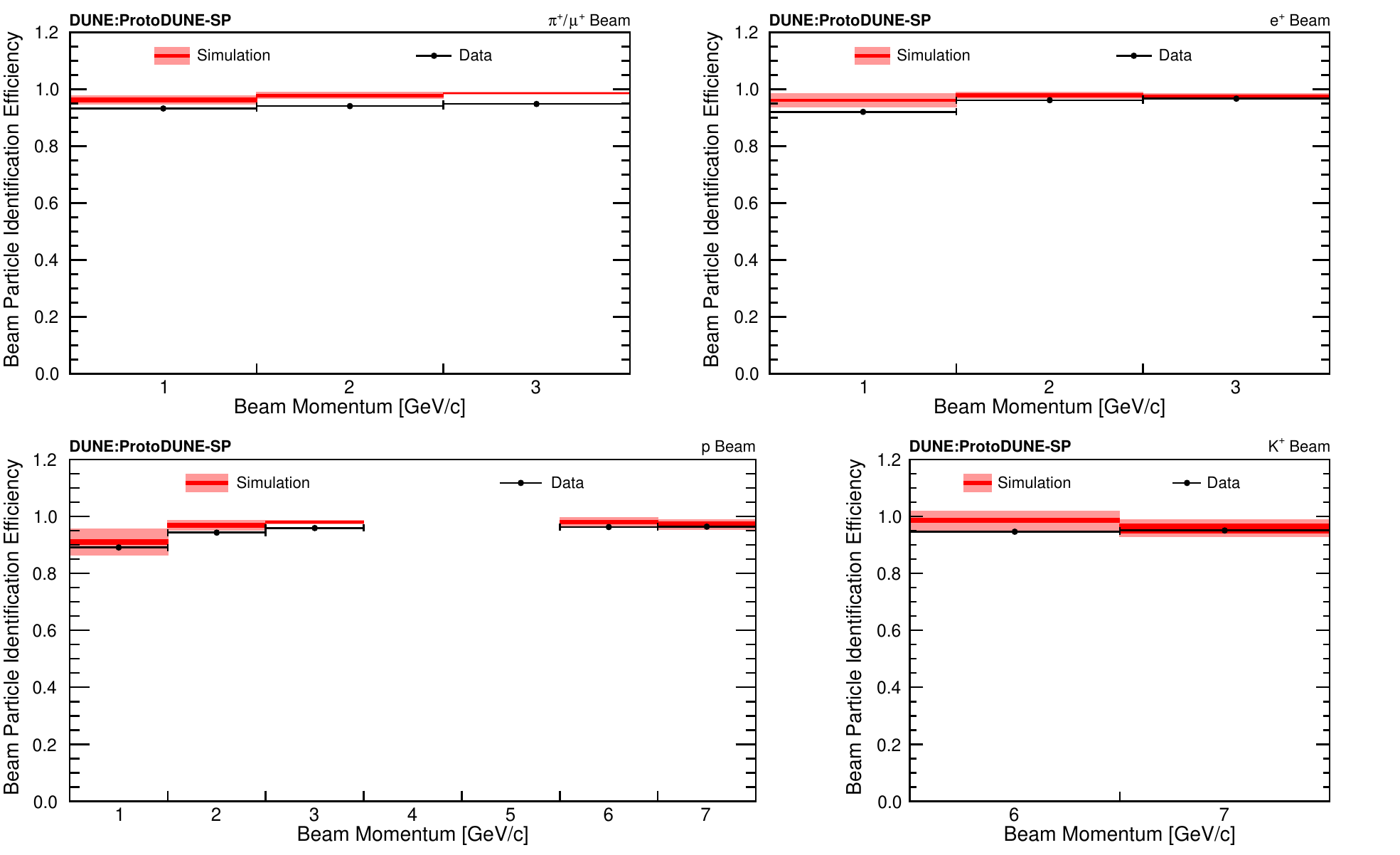}
\caption{The triggered test-beam particle reconstruction efficiency for ProtoDUNE-SP as a function of the beam momentum in data and simulation, for charged pions and antimuons (top left), positrons (top right), protons (bottom left) and charged kaons (bottom right). The pale red simulation band shows the total uncertainty and the statistical-only uncertainty is shown in dark red.}
\label{fig:datamcrecoeff}
\end{figure*}

Figure~\ref{fig:datamcrecoeff} shows the reconstruction and identification efficiency for triggered test-beam charged pions, positrons, protons and kaons as a function of the beam momentum setting. The simulation is shown with statistical uncertainties (darker red) and the quadrature sum of statistical and systematic uncertainties (pale red).

Agreement is seen between the data and simulation and any discrepancies are within 5\%. As expected, the lowest efficiency is seen at 1\,GeV/$c$ where the particles deposit the least energy in the detector. In particular, 1\,GeV/$c$ protons are the most difficult to reconstruct since they travel the shortest distance within the detector.

Figure~\ref{fig:datamcrecoeff} also shows that the fraction of events with a reconstructed beam slice for the pion/muon sample is slightly overestimated in MC compared to data. A number of factors could cause this behaviour. The SCE is underestimated in the simulation~\cite{proto_performance}, which means that the reconstructed beam slices are slightly easier to identify in the simulation. Furthermore, the ratio of muons to pions can differ between data and MC but it is not possible to distinguish these two types of particles due to their similar masses, and a difference in the reconstruction efficiency between pions and muons will induce a data vs MC discrepancy. Finally, in the experimental data, there is a higher probability that beam pion and muon tracks will be broken at the boundary between the first two APAs due to the presence a malfunctioning electron diverter~\cite{proto_performance} that distorts the electric field, which could cause a small reduction in efficiency. Newer versions of the simulation will account for these three effects. However, small differences in the integrated efficiency between data and MC do not have a big impact for the cross-section analyses because the overall normalisation cancels in the cross-section calculation.

\section{Conclusions}
\label{sec:conclusion}
A summary of Pandora, a pattern-recognition software package, has been presented alongside relevant modifications allowing it to be applied to simulated and experimental data interactions in the ProtoDUNE-SP LArTPC detector. Pandora is the primary event reconstruction used in all ProtoDUNE-SP physics analyses and enables the measurement of hadronic cross sections on liquid argon, the primary physics goal of the experiment. The performance of Pandora has been extensively evaluated for simulated charged test-beam and cosmic-ray interactions in the ProtoDUNE-SP detector. Several pattern-recognition metrics have been evaluated enabling a comparison of data and simulation. It is not trivial to extrapolate the performance measured in ProtoDUNE-SP due to the much higher detector occupancy compared to the FD, but it will provide a lower bound on the expected performance and the results presented here demonstrate the potential of Pandora to provide accurate and efficient event reconstruction.

The efficiency to reconstruct the triggered test-beam particle and correctly identify it as of beam origin exceeds 80\% for the majority of particle types ($e^+$, $\pi^+$, $p$, $K^+$, $\mu^+$) and momentum setting combinations (1, 2, 3, 6 and 7\,GeV/$c$). It was also shown that the main cause of these inefficiencies arises from background contamination from both cosmic and beam-halo sources. In background-removed simulation samples the triggered test-beam particle reconstruction and identification efficiency above a few hundred hits is almost 100\% in all cases. A comparison of data and MC shows agreement within 5\% for the reconstruction and identification efficiency for different triggered beam-particle species across the beam momentum values, and possible sources for the small observed efficiency differences were discussed.

Over the coming years, developments to the pattern-recognition are expected from the introduction of new algorithms and the incorporation of deep-learning techniques to drive some key decisions within the Pandora multi-algorithm approach. Examples include improved vertex finding, event slicing and hit classification. Whilst many of these new algorithms are being developed for the DUNE FD, they will be tested on the ProtoDUNE-SP simulation and data. ProtoDUNE-SP has been dismantled but a very similar upgraded detector called ProtoDUNE-HD is under construction in the same cryostat, which is due to commence data taking in 2023.

\begin{acknowledgements}




%
%
The ProtoDUNE-SP detector was constructed and operated on the CERN Neutrino Platform.
We gratefully acknowledge the support of the CERN management, and the
CERN EP, BE, TE, EN and IT Departments for NP04/Proto\-DUNE-SP.
%
%
This document was prepared by the DUNE collaboration using the
resources of the Fermi National Accelerator Laboratory 
(Fermilab), a U.S. Department of Energy, Office of Science, 
HEP User Facility. Fermilab is managed by Fermi Research Alliance, 
LLC (FRA), acting under Contract No. DE-AC02-07CH11359.
%
%
This work was supported by
CNPq,
FAPERJ,
FAPEG and 
FAPESP,                         Brazil;
CFI, 
IPP and 
NSERC,                          Canada;
CERN;
M\v{S}MT,                       Czech Republic;
ERDF, 
H2020-EU and 
MSCA,                           European Union;
CNRS/IN2P3 and
CEA,                            France;
INFN,                           Italy;
FCT,                            Portugal;
NRF,                            South Korea;
CAM, 
Fundaci\'{o}n ``La Caixa'',
Junta de Andaluc\'ia-FEDER,
MICINN, and
Xunta de Galicia,               Spain;
SERI and 
SNSF,                           Switzerland;
T\"UB\.ITAK,                    Turkey;
The Royal Society and 
UKRI/STFC,                      United Kingdom;
DOE and 
NSF,                            United States of America.
%
%
This research used resources of the 
National Energy Research Scientific Computing Center (NERSC), 
a U.S. Department of Energy Office of Science User Facility 
operated under Contract No. DE-AC02-05CH11231.
%

\end{acknowledgements}

\bibliographystyle{unsrtnat}
\bibliography{refs.bib}

\end{document}